\documentclass[aps,prd,twocolumn,preprintnumbers,superscriptaddress,nofootinbib,floatfix]{revtex4}
\usepackage[a4paper, hdivide={1.91cm,,1.165cm}, vdivide={1.83cm,,3.6cm}]{geometry}

\usepackage{amstext,amssymb,amsmath}
\usepackage{amsmath}
\usepackage{graphicx}
\usepackage{xspace}
\usepackage{color}
\usepackage{units}
\usepackage{slashed}
\usepackage{multirow} 
\usepackage{bm} % bold math
\usepackage{ulem}
\usepackage[hyperfootnotes=false]{hyperref}

\begin{document}
\title{Leptonic CP asymmetry and heavy neutrino searches in seesaw scenario}

\author{Arindam Das}
\email{adas@particle.sci.hokudai.ac.jp}
\affiliation{Institute for the Advancement of Higher Education, Hokkaido University, Sapporo 060-0817, Japan}

\author{Wei Liu}
\email{wei.liu@njust.edu.cn}
\affiliation{Department of Applied Physics and MIIT Key Laboratory of Semiconductor Microstructure and Quantum Sensing, Nanjing University of Science and Technology, Nanjing 210094, China}

\author{Supriya Senapati}
\email{ssenapati@njust.ac.cn}
\affiliation{Department of Applied Physics and MIIT Key Laboratory of Semiconductor Microstructure and Quantum Sensing, Nanjing University of Science and Technology, Nanjing 210094, China}

\begin{abstract}
We investigate the prospects for probing heavy Majorana neutrinos in the type-I seesaw framework at the 14 TeV LHC. In this scenario, the light-heavy neutrino mixing enables the production of heavy neutrinos in association with charged leptons, followed by decays into dilepton plus dijet final states. We perform a detector-level cut-based analysis of both same-sign (SS) and opposite-sign (OS) dilepton channels and investigate the sensitivity as a function of the ratio $R_{\ell \ell}$ of SS to OS events. We derive projected constraints on the light-heavy neutrino mixing as a function of the heavy-neutrino mass. For $R_{\ell \ell} \simeq 1$, the sensitivity at $140~\rm{fb}^{-1}$ improves upon current LHC bounds by about one order of magnitude for $M_N \simeq 60-80~\rm{GeV}$, with a further order of magnitude improvement expected at the HL-LHC with $3~\rm{ab}^{-1}$. The sensitivity decreases substantially for smaller $R_{\ell\ell}$. We show that, for $50~\rm{GeV} \lesssim M_N \lesssim 100~\rm{GeV}$, the collider reach is strongly correlated with $R_{\ell\ell}$ and the associated CP asymmetry, providing additional motivation for future measurements of $R_{\ell\ell}$ and heavy-neutrino oscillations at colliders.
\end{abstract}

\maketitle

\section{Introduction}
\label{sec:Intro}

Observations of neutrino oscillation and flavor mixing~\cite{ParticleDataGroup:2024cfk}
have established that Standard Model (SM) neutrinos possess tiny masses. However, the origin of these masses remains unknown, as the SM cannot accommodate them in a natural way. Among the various proposed mechanisms, the seesaw framework~\cite{Yanagida:1980xy, Minkowski:1977sc, Yanagida:1979as, Gell-Mann:1979vob, Glashow:1979nm, Mohapatra:1979ia, Schechter:1980gr} provides one of the simplest and most compelling explanations for their generation. In this scenario, heavy SM-singlet Majorana right-handed neutrinos (RHNs) induce the dimension-five Weinberg operator~\cite{Weinberg:1979sa}, which, after electroweak symmetry breaking, yields naturally suppressed light Majorana neutrino masses. Since the RHNs are singlets under the SM gauge group, they do not interact directly with SM gauge bosons; however, they acquire effective interactions through light-heavy neutrino mixing. Using the general parametrization proposed in Ref.~\cite{Casas:2001sr}, the neutrino Yukawa structure can be constructed to reproduce neutrino oscillation data while remaining consistent with constraints from non-unitarity~\cite{Antusch:2006vwa, Abada:2007ux} and charged lepton flavor violation searches~\cite{MEG:2011naj, BaBar:2009hkt, SuperB:2010cqs}.

In addition to the long-standing puzzle of the origin of tiny neutrino masses, cosmological observations have revealed another fundamental mystery: the baryon asymmetry of the Universe. The observed baryon asymmetry is measured as
\begin{equation}
    Y_B = \frac{n_{B}-n_{\bar{B}}}{s} \simeq 8.7 \times 10^{-11}
\end{equation}
with $\mathcal{O}(10\%)$ precision~\cite{WMAP:2010qai}. While electroweak baryogenesis provides a possible explanation within the Standard Model (SM), it requires a strong first-order electroweak phase transition~\cite{Cohen:1993nk, Funakubo:1996dw, Trodden:1998ym}, which is ruled out by the discovery of the $125~\mathrm{GeV}$ Higgs boson~\cite{CMS:2012qbp, ATLAS:2012yve}. This motivates scenarios beyond the SM, among which leptogenesis offers an appealing mechanism that links the origin of neutrino masses and the baryon asymmetry~\cite{Fukugita:1986hr}. In this framework, a lepton asymmetry is generated from out of equilibrium decays of heavy Majorana neutrinos and subsequently converted into a baryon asymmetry via sphaleron processes that violate $B+L$ while conserving $B-L$, leading to $Y_B = -\frac{27}{79} Y_L$ in the SM~\cite{Manton:1983nd, Klinkhamer:1984di, Khlebnikov:1988sr}.

The same framework that accounts for the smallness of neutrino masses and the baryon asymmetry via leptogenesis  predicts the existence of heavy RHNs, making their direct experimental search an important target.
If sterile Majorana RHNs possess masses around the electroweak scale or above, they can be directly produced at high-energy colliders through various initial-state processes, depending on the nature of the colliders~\cite{Pilaftsis:1997dr, Bray:2007ru, Deppisch:2015qwa, Das:2015toa, Das:2016hof, Bolton:2019pcu}. In standard type-I seesaw scenarios, the heavy Majorana RHNs mix with the light active neutrinos of the SM. As a consequence of this light-heavy neutrino mixing, the heavy neutrinos acquire effective interactions with the SM gauge bosons, thereby allowing their production and detection at collider experiments. 

However, in a standard type-I seesaw, the light-heavy mixing angles large enough to give observable collider production typically require sizable Yukawa couplings, which in turn would generate light neutrino masses that are far too large unless delicate cancellations occur. This creates a tension between achieving collider-accessible heavy neutrinos and maintaining sub-eV light neutrino masses in the absence of a protective symmetry~\cite{Kersten:2007vk, Drewes:2019byd}. This tension is naturally resolved if the heavy neutrino sector approximately respects a generalized lepton number symmetry, which ensures that the smallness of the light neutrino masses is technically natural rather than the result of fine-tuned cancellations. In this symmetry limit, lepton number violating effects are suppressed relative to lepton number conserving interactions, and the residual violation is conveniently quantified by $R_{\ell\ell}$, which encodes the observable impact of the small symmetry-breaking mass splitting in dilepton final states~\cite{Asaka:2005pn, Canetti:2012kh, Roy:2010xq, Drewes:2019byd}.

In this paper we consider the seesaw scenario where two generations of the Majorana type heavy neutrinos belong to the GeV scale being nearly degenerate. The third generation of the heavy neutrino is very heavy with a mass of $\mathcal{O}(10^{13})$ GeV. Decay of heavy neutrino following $N \to \ell H$ mode takes place through tree level and 1-loop quantum level processes~\cite{Liu:1993tg} through the Yukawa interaction and their interference create a CP asymmetry in the lepton sector which play a pivotal role to understand the origin of the Baryon Asymmetry of the Universe (BAU)~\cite{WMAP:2010qai}. While the heavy neutrinos are nearly degenerate in nature, the CP asymmetry can be substantially enhanced~\cite{Flanz:1994yx, Flanz:1996fb, Covi:1996fm, Pilaftsis:1997jf, Pilaftsis:2003gt}. This mechanism is known as the resonant leptogenesis. Applying general parametrization~\cite{Casas:2001sr} of the Dirac Yukawa coupling among the SM-singlet RHN, SM lepton and Higgs doublets, reproducing the neutrino oscillation data we formulate the leptonic CP asymmetry in terms of the small mass difference between the two nearly degenerate heavy neutrinos and their corresponding decay widths.

We study the production of heavy Majorana neutrinos at the Large Hadron Collider (LHC) with center of mass energy $\sqrt{s}=14$ TeV. These heavy neutrinos are predominantly produced in association with a charged lepton through an $s$-channel process mediated by a $W$ boson. After production, we consider their dominant semileptonic decay mode into a charged lepton and two jets, which leads to a characteristic final state with same-sign (SS) and opposite-sign (OS) dileptons plus two jets. To quantify these signatures, we consider the ratio $R_{\ell\ell}$. In the limit of exact lepton number conservation, the SS and OS dilepton rates become equal, corresponding to $R_{\ell\ell}\to 1$. Deviations from this limit arise due to the small mass splitting, which also governs the size of lepton number violation and can be connected to the leptonic CP asymmetry in resonant leptogenesis. We perform a scan of the leptonic CP asymmetry as a function of $R_{\ell\ell}$ over the range $0 < R_{\ell\ell} \leq 1$, imposing constraints from neutrino oscillation data and varying the Dirac and Majorana CP phases. For the collider analysis to compute the signal and corresponding SM backgrounds we simulate events using {\tt MadGraph5}~\cite{Alwall:2011uj} followed by {\tt PYTHIA8.3}~\cite{Bierlich:2022pfr} for hadronization and {\tt Delphes}~\cite{deFavereau:2013fsa} for detector simulation CMS card. We use the NNPDF~\cite{NNPDF:2014otw} parton distribution function (PDF) for our analysis. 

The paper is organized as follows. In Sec.~\ref{sec:Model}, we present the seesaw model. In Sec.~\ref{sec:NProd}, we discuss heavy neutrino production and  dilepton signatures. In Sec.~\ref{sec:CPAsymmetry}, we study leptonic CP asymmetry and its connection to the ratio of SS and OS dilepton events. Collider searches for SS and OS dileptons with two jets at the LHC are discussed in Sec.~\ref{sec:Collider}. The heavy-neutrino decay widths used in our analysis are summarized in Appendix~\ref{app:Ndecays}. We conclude in Sec.~\ref{sec:Conclusion}.

\section{Model} 
\label{sec:Model}

We consider the type-I seesaw framework obtained by extending the SM with three generations of gauge-singlet right-handed neutrinos (RHNs), $N_R^\beta$ ($\beta=1,2,3$). The relevant Yukawa and Majorana mass terms are
\begin{equation}
\mathcal{L} \supset -Y^{\alpha\beta} \overline{\ell_L^{\alpha}} \widetilde H N_R^{\beta} -\frac{1}{2} M_N^{\alpha \beta} \overline{N_R^{\alpha C}} N_R^{\beta} + \rm{H. c.} .
\label{eq:typeI}
\end{equation}
where $\ell_L^\alpha$ denotes the SM lepton doublet, $\widetilde H = i\sigma_2 H^\ast$, and $M_N$ is the Majorana mass matrix of the RHNs. After the spontaneous electroweak symmetry breaking, the Higgs field acquires a vacuum expectation value (VEV), $\langle H\rangle = v/\sqrt{2}$ with $v=246$ GeV, generating the Dirac neutrino mass matrix, $m_D=\frac{v}{\sqrt2}Y$. In the basis $(\nu_L,N_R^C)$, the full neutrino mass matrix is
\begin{equation}
m_{\nu} = 
\begin{pmatrix}
0 & m_{D} \\
m_{D}^{T} & M_{N}
\end{pmatrix}.
\label{eq:typeInu}
\end{equation}
For $M_N\gg m_D$, diagonalizing the matrix given in Eq.~\eqref{eq:typeInu} we obtain the seesaw formula for the light Majorana neutrinos as
\begin{equation}
m_{\nu} \simeq - m_{D} M_{N}^{-1} m_{D}^{T}.
\label{eq:nuMass}
\end{equation}
For a representative heavy-neutrino mass scale $M_N \sim \mathcal{O}(100)$ GeV, the canonical seesaw relation in Eq.~\eqref{eq:nuMass} implies Yukawa couplings of order $Y \sim \mathcal{O}(10^{-6})$ in order to reproduce light-neutrino masses $m_\nu \sim \mathcal{O}(0.1)$ eV. Such small couplings lead to highly suppressed light-heavy neutrino mixing and consequently negligible heavy-neutrino production rates at colliders. However, in the generalized Casas-Ibarra parametrization a larger Yukawa couplings can be realized while remaining consistent with neutrino oscillation data. This allows enhanced light-heavy neutrino mixing and produces heavy Majorana neutrino at the high energy colliders with a sizable cross section.

Since the RHNs are singlets under the SM gauge group, they do not possess direct gauge interactions with the SM fields. After electroweak symmetry breaking, however, the Dirac mass term induces mixing between the light and heavy neutrino sectors. Consequently, the flavor eigenstates of the light neutrinos can be expressed in terms of the light ($\nu_i$) and heavy ($N_i$) mass eigenstates as
\begin{equation}
\nu_\alpha \simeq  {(U_{\rm PMNS})}_{\alpha i} \nu_i  + V_{\alpha i} N_i,
\label{eq:Flavor2Mass}
\end{equation}
where $V_{\alpha i}(= m_D/M_{N_i})$ is the mixing between light and heavy mass eigenstates. Although suppressed by the seesaw hierarchy, this mixing provides the primary production and decay mechanism of heavy neutrinos at colliders. The mixing matrix, ${U_{\rm{PMNS}}}$, denotes the Pontecorvo-Maki-Nakagawa-Sakata (PMNS) matrix which diagonalizes the light neutrino mass matrix as
\begin{equation}
U_{\rm PMNS}^T m_{\nu} U_{\rm PMNS} = D_m = \rm{diag} (m_1, m_2, m_3)
\end{equation}
with the standard decomposition $U_{\rm PMNS} = V(\theta_{12}, \theta_{13}, \theta_{23}, \delta) . P(\alpha_1, \alpha_2)$ where $V$ contains the three neutrino mixing angles and the Dirac CP-violating phase $\delta$,
\begin{equation}
\resizebox{\columnwidth}{!}{$
V =
\begin{pmatrix}
c_{12} c_{13} &
s_{12} c_{13} &
s_{13} e^{-i\delta} \\
- s_{12} c_{23} - c_{12} s_{23} s_{13} e^{i\delta} &
c_{12} c_{23} - s_{12} s_{23} s_{13} e^{i\delta} &
s_{23} c_{13} \\
s_{12} s_{23} - c_{12} c_{23} s_{13} e^{i\delta} &
- c_{12} s_{23} - s_{12} c_{23} s_{13} e^{i\delta} &
c_{23} c_{13}
\end{pmatrix}
$}
\end{equation}
with $c_{ij} \equiv \cos\theta_{ij}$ and $s_{ij} \equiv \sin\theta_{ij}$ while $P(\alpha_1, \alpha_2) = \rm{diag} \left(1, e^{i\alpha_1/2}, e^{i\alpha_2/2}\right)$ where $\alpha_{1,2}$ are Majorana CP phases.

To establish the connection between the neutrino Yukawa sector and low-energy neutrino observables, we employ the Casas-Ibarra parametrization, in which the Dirac Yukawa matrix is expressed as
\begin{equation}
Y = \frac{\sqrt{2}}{v} U_{\rm PMNS} \sqrt{D_m} \mathcal{R}  \sqrt{D_M},
\end{equation}
where $v=246$ GeV is the electroweak VEV. Considering heavy neutrinos in diagonal basis, we write $D_M = \rm{diag} (M_1, M_2, M_3)$ and $\mathcal{R}$ is a general complex orthogonal matrix satisfying $\mathcal{R} \mathcal{R}^T = \mathbb{I}$. It can be parametrized as
\begin{equation}
\mathcal{R} = \begin{pmatrix}
\cos(a+ib) & \sin(a+ib) & 0 \\
- \sin(a+ib) & \cos(a+ib) & 0 \\
0 & 0 & 1
\end{pmatrix},
\end{equation}
with $a,~b \in \mathcal{R}$. The structure of the light-neutrino mass matrix, $D_m$, depends on the neutrino mass ordering. For normal ordering (NO), corresponding to $m_1<m_2<m_3$, and inverted ordering (IO), corresponding to $m_3<m_1<m_2$, the diagonal mass matrices can be written in terms of the measured mass-squared differences as
\begin{align}
& D_m^{\rm NO}
=\begin{pmatrix}
m_1 & 0 & 0 \\
0 & \sqrt{m_1^2+ \Delta m_{21}^2} & 0 \\
0 & 0 & \sqrt{m_1^2+ \Delta m_{31}^2}
\end{pmatrix}, \nonumber \\
& D_m^{\rm IO}
=\begin{pmatrix}
\sqrt{m_3^2+|\Delta m_{32}^2|-\Delta m_{21}^2} & 0 & 0 \\
0 & \sqrt{m_3^2+ |\Delta m_{32}^2}| & 0 \\
0 & 0 & m_3
\end{pmatrix},
\end{align}
while the form of $U_{\rm PMNS}$ is identical in both cases. The choice of mass ordering determines which light neutrino mass eigenstates dominantly control the CP asymmetry through the factors $m_i$ appearing in the Casas-Ibarra parametrization.

Due to the mixing between the light and heavy neutrinos, the flavor eigenstates can be expressed in terms of the light and heavy mass eigenstates according to Eq.~\eqref{eq:Flavor2Mass}. Consequently, the charged-current (CC) and neutral-current (NC) interactions involving neutrinos can be written in the mass basis. The CC interaction is given by
\begin{equation}
\mathcal{L}_{\rm CC} \supset -\frac{g}{\sqrt{2}} W_{\mu} \overline{\ell_\alpha} \gamma^{\mu} P_L V_{\alpha i} N_i  + {\rm H.c.}, 
\label{eq:CC}
\end{equation}
where $\ell$ denotes three generations of the charged leptons in vector form. The corresponding NC interaction is
\begin{equation}
\resizebox{\columnwidth}{!}{$
\mathcal{L}_{\rm NC} \supset -\frac{g}{2 \cos\theta_{\rm W}}  Z_{\mu} \left[ \overline{N_i} \gamma^{\mu} P_L |V_{\alpha i}|^2 N_i + ( \overline{\nu_\alpha} \gamma^{\mu} P_L V_{\alpha i}  N_i + \rm{H.c.} ) \right], 
\label{eq:NC}
$}
\end{equation}
where $\theta_{\rm W}$ denotes the Weinberg mixing angle with $\sin^2 \theta_W = 0.2229$ and $P_L = \frac{1}{2} (1- \gamma_5)$ is the projection operator. The heavy-neutrino decay widths used in this analysis are summarized in
Appendix~\ref{app:Ndecays}. For heavy Majorana neutrinos, the decay channels related by charge conjugation contribute equally to the total width. In particular, the decay modes given in Eqs.~\eqref{eq:nul1l2},~\eqref{eq:l1p},~\eqref{eq:l1v},~and~\eqref{eq:l1ud} receive identical contributions from their charge-conjugate processes, effectively doubling the corresponding partial widths. Such contributions are absent in the Dirac-neutrino case. The total decay width of a Majorana heavy neutrino can therefore be written as
\begin{align}
\Gamma_N  = & \sum_{\ell_1, \ell_2 (\ell_1\neq \ell_2)} \left( 2\Gamma(N \to \ell_1^- \ell_2^+\nu_{\ell_2}) + \Gamma(N \to \nu_{\ell_1} \ell_2^- \ell_2^+) \right) \nonumber \\
& + \sum_{\ell_2} \Gamma(N \to \nu_{\ell_2} \ell_2^- \ell_2^+) + \sum_{\ell_1}\Gamma(N \to \nu_{\ell_1}\nu\bar{\nu}) \nonumber \\
& + \Gamma^{\rm semilepton},
\end{align}
where the semileptonic contribution from the heavy neutrinos is given by
\begin{align}
 \Gamma^{\rm semilepton} & = \theta(\mu_0-M_N) \sum_{\ell_1, P, V} \Big[ 2\Gamma(N \to \ell_1^-P^+) \nonumber \\
 &  + 2\Gamma(N \to \ell_1^-V^+) + \Gamma(N \to \nu_{\ell_1}P^0) \nonumber \\
 & + \Gamma(N \to \nu_{\ell_1}V^0)\Big] + \theta(M_N - \mu_0) \nonumber \\
 & \sum_{\ell_1,q,q'}[\Gamma(N \to \nu_{\ell_1} q\bar{q}) + 2\Gamma(N \to \ell_1^- q\bar{q'})]
\end{align}
Here, $\mu_0$ denotes the mass threshold for which we adopt semileptonic contributions via quark-antiquark production, and we consider this as $\mu_0=957.8$ MeV. For the limit $M_N <\mu_0$ heavy neutrinos become kinematically allowed to decay into light mesons like $\pi^{0, \pm}$, $\rho^{0,\pm}$, $\eta$, $\omega$, and strange mesons including $K^{0, \pm}$ and $K^{*0,\pm}$. In contrast, for mass range $M_N > \mu_0$, we can consider quarks as degrees of freedom, and then the semileptonic decay will consist of quark pairs such as $q\bar{q}$ and $q\bar{q}^\prime$, respectively. 

\begin{figure}[htb!]
    \centering
\includegraphics[width=1.0\linewidth]{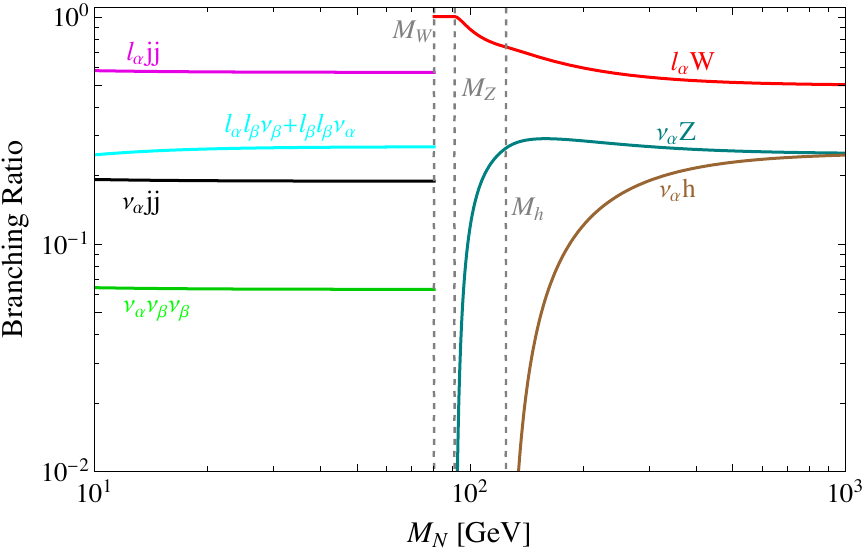}
\caption{Branching ratios of the RHN $N$ as a function of its mass $M_N$. The dashed vertical lines denote the kinematic thresholds corresponding to $M_W$, $M_Z$, and $M_h$. Below the electroweak scale, three-body decay channels dominate, while above these thresholds the two-body modes involving on-shell $W$, $Z$, and Higgs bosons become accessible and gradually dominate the decay pattern.}
\label{fig:Branching_Ratios}
\end{figure}

The branching ratios of the RHNs as a function of its mass has been shown in Fig.~\ref{fig:Branching_Ratios}.
For $M_N < M_W$, the decays proceed exclusively through off-shell weak bosons, leading to nearly mass-independent branching fractions for the three-body final states. Once the $W$-boson threshold is crossed, the channel $N \to \ell_\alpha W$ opens and rapidly becomes dominant. The subsequent opening of the $N \to \nu_\alpha Z$ and $N \to \nu_\alpha h$ channels at $M_Z$ and $M_h$, respectively, redistributes the total decay width among the available two-body modes. The corresponding three-body decay channels become increasingly suppressed at large $M_N$.

\section{Heavy Neutrino Production and Dilepton Signatures}
\label{sec:NProd}

At the LHC with 14 TeV center of mass energy, Majorana type heavy neutrinos can be produced through the CC interactions by the $W$ boson exchange in $s$-channel in association with a charged lepton. The dominant production process of heavy neutrinos at the parton level is $u \bar{d}\rightarrow \ell^{+} N$ (and $\bar{u} d \rightarrow \ell^{-} N$). The differential scattering cross section of this process is given by
\begin{align}
\frac{d \hat{\sigma}_{\rm{LHC}}}{d \cos\theta} = & 
(3.89 \times 10^8 ~{\rm pb}) \times \frac{\beta}{32 \pi \hat{s}} \frac{\hat{s} +M_N^2}{\hat s} \left( \frac{1}{2} \right)^2 3 \left( \frac{1}{3} \right)^2 \nonumber \\
 & \times 
  \left|V_{\ell N} \right|^2  \frac{g^4}{4} 
\frac{({\hat s}^2-M_N^4)(2 + \beta \cos^2 \theta)}
{({\hat s} - M_W^2)^2+ M_W^2 \Gamma_W^2},
\end{align}
where $\sqrt{\hat s}$ denotes the center of mass energy of the colliding partons, $M_N$ is the mass of heavy neutrino, and $\beta =({\hat s}-M_N^2)/({\hat s}+M_N^2)$. The total production cross section of RHNs in association with a charged lepton at the LHC reads as
\begin{align}
\sigma_{\rm{LHC}} & =
\int d \sqrt{\hat s} \int d \cos \theta 
\int^1_{{\hat s}/E_{\rm{CM}}^2} dx~ 
\frac{4 {\hat s}}{x E_{\rm{CM}}^2} 
f_u(x,Q) \nonumber \\
& \times f_{\bar d}\left( \frac{\hat s}{x E_{\rm{CM}}^2},Q \right)  
\frac{d \hat{\sigma}_{\rm{LHC}}}{d \cos\theta} + (u \to {\bar u}, {\bar d} \to d), 
\label{eq:xLHC}
\end{align}
taking $E_{\rm{CM}} = 14$ TeV as the center of mass energy of the LHC. In the numerical analysis we employ NNPDF~\cite{NNPDF:2014otw}, for the parton distribution functions for up$(u)$-quark ($f_u$) and down $({\bar d)}$-quark ($f_{\bar d}$). Here $Q$ denotes the factorization scale and it is fixed at $\sqrt{\hat s}$. In our analysis we consider dominant decay mode of heavy neutrinos following $N \to \ell jj$ through $W$ boson showing same sign and opposite sign signature in association with two jets. As a result we can define a key observable in terms of the ratio of SS and OS dilepton events
\begin{equation}
R_{\ell \ell}=\frac{N_{\rm SS}}{N_{\rm OS}}
\end{equation}
where $N_{\rm SS (OS)}$ is the number of SS (OS) events. In this analysis, to calculate events we consider 140 fb$^{-1}$ and a futuristic 3 ab$^{-1}$ luminosities in LHC. Considering the physical mass splitting between the heavy neutrino states producing SS and OS di-leptons $(\Delta M)$ following~\cite{Anamiati:2016uxp, Das:2017hmg, Drewes:2019byd} we define
\begin{equation}
R_{\ell \ell}=\frac{\Delta M^2}{2\Gamma^2+ \Delta M^2}
\label{eq:Rll}
\end{equation}
where $\Gamma$ represents the total decay width of heavy neutrinos. There are two limiting cases when $R_{\ell \ell} \to 1$ as $\Gamma/ \Delta M \to 0$
resulting the Majorana nature of the heavy neutrinos whereas $R_{\ell \ell} \to 0$ as $\Delta M/\Gamma \to 0$ resulting the Dirac nature of the heavy neutrinos, respectively.

\section{Connection of \texorpdfstring{$R_{\ell\ell}$}{Rll} to CP asymmetry}
\label{sec:CPAsymmetry}

\begin{figure*}[htb!]
\centering
\includegraphics[width=0.45\linewidth]{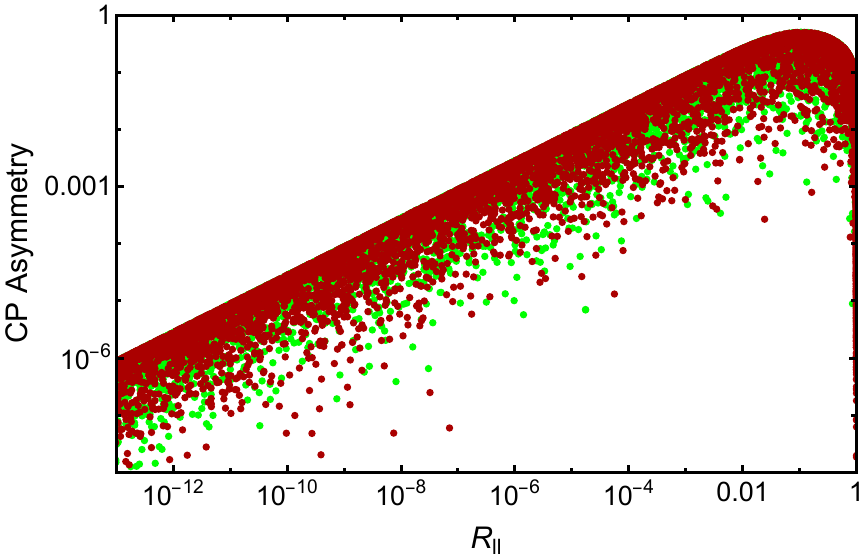} \qquad 
\includegraphics[width=0.45\linewidth]{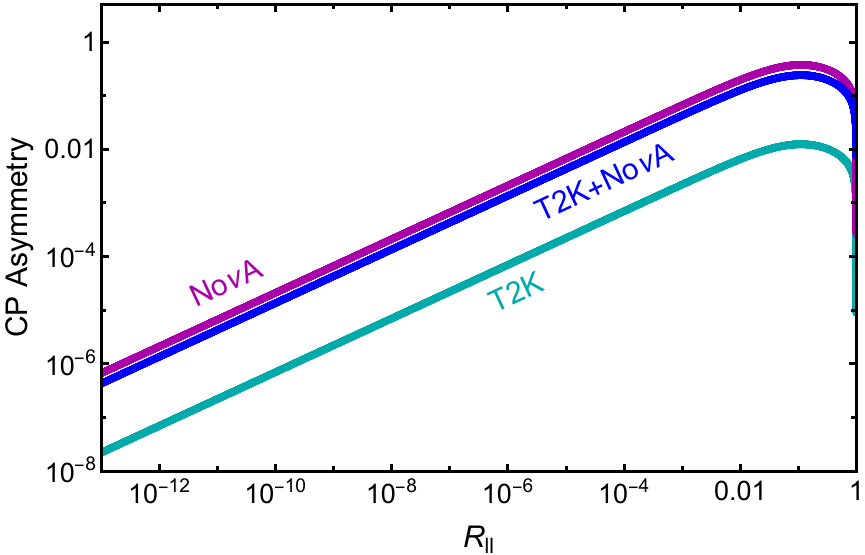}
\caption{Absolute value of the CP asymmetry parameter as a function of $R_{\ell\ell}$. Left panel: Results of a random scan over the model parameter space for the NO (green) and IO (red), including variations of the light and heavy neutrino masses, the complex orthogonal matrix $R$, and the CP-violating phases. Right panel: Benchmark predictions obtained using the best-fit Dirac CP phases from T2K, NO$\nu$A, and the combined T2K+NO$\nu$A analyses.}
\label{fig:EpsilonRll}
\end{figure*}

If the heavy neutrinos $N_i$ and $N_j$ are nearly degenerate, the CP asymmetry in the decay of a heavy Majorana neutrino into leptons is resonantly enhanced which is a prerequisite of resonant leptogenesis. The flavor independent CP asymmetry is given by~\cite{Chen:2007fv}
\begin{align}
\epsilon_i \simeq \sum_{j \ne i} 
\frac{\operatorname{Im}[(Y Y^\dagger)_{ij}^2]}{(Y Y^\dagger)_{ii} (Y Y^\dagger)_{jj}} \cdot 
\frac{(M_{N_i}^2 - M_{N_j}^2) M_i \Gamma_j}
{(M_{N_i}^2 - M_{N_j}^2)^2 + M_i^2 \Gamma_j^2}
\label{eq:Epsilon1}
\end{align}
where $M_{i,j}$ are the masses of $i$th and $j$th heavy neutrinos and $\Gamma_j$ is the total decay width of the $j$th heavy neutrino, $\Gamma_j = \frac{1}{8\pi} (Y Y^\dagger)_{jj} M_{j}$. In the limit $\Delta M_{ij} = M_{N_i} - M_{N_j} \ll M_{N_i}$, the Eq.~\eqref{eq:Epsilon1} becomes
\begin{align}
\epsilon_i \simeq \frac{1}{2} \frac{\operatorname{Im}[(YY^\dagger)^2_{ij}]}{(YY^\dagger)_{ii} (YY^\dagger)_{jj}} \frac{\Delta M_{ij} \Gamma_j}{(\Delta M_{ij})^2+\Gamma_j^2/4}
\label{eq:Epsilon2}
\end{align}
exhibiting the fact that resonant enhancement occurs when $\Delta M_{ij} \simeq \Gamma_j/2$, allowing a large CP asymmetry for low-scale heavy neutrinos. From Eq.~\eqref{eq:Rll} we obtain $\Delta M_{ij}= \sqrt{\frac{2 R_{\ell \ell}}{1 - R_{\ell\ell}}} \Gamma_j$ and using this in Eq.~\eqref{eq:Epsilon2} we obtain
\begin{align}
\epsilon_i \simeq \frac{\operatorname{Im}[(YY^\dagger)^2_{ij}]}{(YY^\dagger)_{ii} (YY^\dagger)_{jj}} \frac{2 \sqrt{ 2 R_{\ell \ell} (1-R_{\ell \ell})}}{(1+ 7 R_{\ell \ell})}.
\label{eq:RllEpsilon}
\end{align}
The above equation is valid only for $0 < R_{\ell \ell} <1$ and therefore it breaks down in the limiting cases of pure Dirac $(R_{\ell \ell}=0)$ and pure Majorana $(R_{\ell \ell}=1)$ neutrinos. For $R_{\ell \ell}=0.5$, $\epsilon=\epsilon_{\rm max}$, i.e., it results a maximal CP asymmetry. Now, using Casas-Ibarra parametrization as discussed in Sec.~\ref{sec:Model}, we have shown the correlation between the magnitude of the CP asymmetry and the parameter $R_{\ell\ell}$ in the left panel of Fig.~\ref{fig:EpsilonRll}.
For this scan we have used the the Global fit data at 3$\sigma$ from NuFit:6.0~\cite{Esteban:2024eli} with the inclusion of Super-Kamiokande data as given in Tab.~\ref{tab:Oscillation_parameter}.
\begin{table}[htb!]
    \centering
    \begin{tabular}{|c|c|c|}
    \hline \hline
    {\bf Parameter} & {\bf NO} & {\bf IO} \\
    \hline \hline
     \rule{0pt}{2.5ex}
    $\theta_{12}\,(^\circ)$ & $33.68^{+2.27}_{-2.05}$ & $33.68^{+2.27}_{-2.05}$ \\
  $\theta_{13}\,(^\circ)$ & $8.56^{+0.33}_{-0.37}$ & $8.59^{+0.34}_{-0.34}$ \\
$\theta_{23}\,(^\circ)$ & $43.3^{+6.6}_{-2.0}$ & $47.9^{+1.9}_{-6.4}$ \\
$\Delta m_{21}^2\,(10^{-5}\,\text{eV}^2)$ & $7.49^{+0.56}_{-0.57}$ & $7.49^{+0.56}_{-0.57}$ \\
$\Delta m_{31(32)}^2\,(10^{-3}\,\text{eV}^2)$ & $2.513^{+0.065}_{-0.062}$ & $-2.484^{+0.063}_{-0.063}$ \\
$\delta\,(^\circ)$ & $212^{+152}_{-88}$ & $274^{+61}_{-73}$ \\
\hline \hline
\end{tabular}
\caption{Best-fit values and $3\sigma$ intervals for the neutrino oscillation parameters adopted in this analysis, obtained from the NuFit~6.0 global fit~\cite{Esteban:2024eli} with the inclusion of Super-Kamiokande data.}
\label{tab:Oscillation_parameter}
\end{table}
The other parameter ranges adopted for this scans are presented in Tab.~\ref{tab:param_scan}. Also, we keep $M_{N_3}$ to be fixed at $10^{13}$ GeV so that it decouples from the theory.
\begin{table}[hbt!]
\centering
\begin{tabular}{lr}
\toprule
\textbf{Parameter} & \textbf{Range} \\
\hline
\textit{General orthogonal matrix $\mathcal{R}$}\\
$a$ & $[-10, 10]$ \\
$b$ & $[-10, 10]$ \\ \\
\hline
\textit{Light neutrinos}\\
$\alpha_1$, $\alpha_2$ [rad] & $[-\pi, \pi]$ \\
$m_{\nu_{\rm min}}$ [eV] & $[10^{-5}, 10^{-2}]$ \\ \\
\hline
\textit{RHNs}\\
$M_{N_j}$ [GeV] & $[50,1000]$ \\
$\Delta M_{12}$ [eV] & $[10^{-2},10^9]$ \\
\hline
\end{tabular}
\caption{Parameter ranges adopted for the scans and we keep $M_{N_3}$ to be fixed at $10^{13}$ GeV so that it decouples from the theory. $\alpha_{1,2}$ are the Majorana phases. Here $m_{\nu_{\rm min}}$ denotes the lightest active neutrino mass and it is a free parameter being denoted by $m_1$ in the NO and $m_3$ in the IO cases, respectively.}
\label{tab:param_scan}
\end{table}
We see that larger values of $R_{\ell\ell}$ generally yielding larger CP asymmetries. The IO tends to populate slightly larger values of CP asymmetry compared with the NO, although substantial overlap exists between the two spectra. The maximal asymmetry approaches $\mathcal{O}(1)$ for $R_{\ell\ell} \sim 0.1$, while it becomes strongly suppressed for smaller values of $R_{\ell\ell}$ as the CP asymmetry in that regime proportional to $R_{\ell\ell}^{1/2}$ due to the term being a function of $R_{\ell \ell}$ in Eq.~\eqref{eq:RllEpsilon}. 

Next, in addition to the full scan over the parameter space, we consider three benchmark points (BPs). The right panel of Fig.~\ref{fig:EpsilonRll} shows the results for the BPs of the Dirac CP phase taken from the central values measured by the T2K~\cite{T2K:2023smv} and NO$\nu$A~\cite{NOvA:2021nfi} experiments as $\delta = -0.5\pi$ and $\delta = 0.8 \pi$, respectively. We also used the combined central value of the Dirac CP phase $\delta = -1.08 \pi$ from T2K and NO$\nu$A~\cite{T2K:2025wet} experiments. The neutrino oscillation parameters are fixed to their best-fit values listed in Tab.~\ref{tab:Oscillation_parameter}. Throughout this analysis, we take $m_{\nu_{\rm min}} = 10^{-3}\,\mathrm{eV}$, $M_{N_1} = 100\,\mathrm{GeV}$, and $\alpha_{1,2} = \pi$, while imposing $a=b$. The value of $b$ is determined by requiring a light-heavy neutrino mixing of $|V_{\alpha N}|^2 \approx 10^{-5}$, corresponding to the phenomenologically relevant parameter region considered in this work. Therefore, for each benchmark value of $\delta$, the prediction is represented by a curve rather than a band or scattered region.
As we see from the right panel of Fig.~\ref{fig:EpsilonRll} that while NO$\nu$A and T2K+NO$\nu$A predictions are nearly identical over the entire $R_{\ell \ell}$ range, whereas the T2K result is consistently smaller by about one to two orders of magnitude. The Dirac CP phase $\delta$, directly affect the interference structure entering the leptogenesis asymmetry. Although for T2K, $\delta \simeq - 0.5 \pi$ corresponds to maximal CP violation in neutrino oscillations, the lepton asymmetry depends on a more complicated combination of PMNS matrix elements through $\rm{Im}[(YY^\dagger_{12})^2]$. The relevant interference terms are therefore not maximized solely by the oscillation measure of CP violation. For the T2K parameter set, cancellations among the PMNS-induced contributions suppress the imaginary part of $\rm(YY^\dagger_{12})$, resulting in significantly smaller values of CP asymmetry. In contrast, the NO$\nu$A and T2K+NO$\nu$A best-fit phases generate very similar interference patterns in the Yukawa sector, leading to comparable values of $\rm{Im}[(YY^\dagger_{12})^2]$. Therefore, their predicted CP asymmetries is almost overlapping.

Before concluding this section, we discuss how the non-unitarity constraints can be used to place an upper bound on the imaginary part of the Casas-Ibarra matrix parameter $(b)$. The non-unitarity constraints considered in this work arise from the mixing between the light neutrinos and the heavy Majorana states introduced in the type-I seesaw mechanism. Since the full neutrino mixing matrix contains both light and heavy degrees of freedom, the effective $3 \times 3$ leptonic mixing matrix describing the light neutrino sector is not exactly unitary. After electroweak symmetry breaking, the neutrino mass matrix is given in Eq.~\eqref{eq:typeInu} and in the seesaw limit, $M_N \gg m_D$, the corresponding light-heavy mixing matrix is approximately $V_{\alpha N} \simeq m_D M_N^{-1}$. As a consequence, the effective leptonic mixing matrix for the light neutrinos reads as $ \mathcal{N} \simeq (1-\eta)U_{\rm PMNS}$, where the non-unitarity parameter is defined by $\eta_{\alpha \beta} = \frac{1}{2}\sum_j V_{\alpha N_{j}} V_{\beta N_{j}}^\dagger$. In the type-I seesaw framework, all amplitudes involving heavy neutrinos are suppressed by $U^2_{\alpha N_{j}} \equiv |V_{\alpha N_{j}}|^2$~\cite{Klaric:2021cpi}. For phenomenological applications, it is convenient to introduce the summed quantity as $U^2 \equiv \sum_{\alpha,j} U^2_{\alpha N_{j}}$ which represents the overall mixing between the heavy and light neutrino sectors. It is particularly relevant for phenomenology, as it controls the strength of RHNs interactions with SM particles. In the seesaw parameterization, it can be expressed as
\begin{equation}
U^2 = \frac{\sum_i  m_i}{M_{N_i}} \cosh(2b).
\label{eq:U2}
\end{equation}
Also, we have $\mathrm{Tr}(\eta)=\frac{1}{2} \mathrm{Tr} \left( V V^\dagger \right) \equiv \frac{1}{2} U^2$. Therefore, one can derive a maximum limit on $b$ for maximum value of $\eta$ as for large value of $b$ it can be approximated as
\begin{equation}
b < \frac{1}{2} \ln \left( \frac{4 \eta_{\max} M_{N_i}}{\sum_i m_i}\right).
\label{eq:Limit_b}
\end{equation}

For representative values relevant to the present study, $ M \sim 100 ~\rm{GeV},~ m_\nu \sim 0.05 ~\rm{eV},~ \eta_{\rm{max}} \sim 10^{-3}$, the resulting constraint is of order $b_{\rm{max}} \sim {\cal O}(10)$. Thus, values of $b$ up to approximately $10$ remain broadly compatible with current non-unitarity bounds, whereas substantially larger values would generally lead to excessive light-heavy mixing. The non-unitary nature of $\mathcal{N}$ leads to strong constraints on its elements from global analyses of neutrino oscillation experiments, precision electroweak measurements involving weak boson decays, and searches for charged-lepton flavor violation~\cite{Antusch:2006vwa, Abada:2007ux, Ibarra:2010xw, Ibarra:2011xn, Dinh:2012bp}. Using the non-unitarity bounds given in Ref.~\cite{Das:2017nvm}, the more stringent flavor-dependent constraints, particularly those involving the $e-\mu$ sector, $|\mathcal{N}\mathcal{N}^\dagger| \ll 1.288 \times 10^{-5}$ and this can further reduce the allowed range to approximately $b \lesssim 7-9$.

\section{Collider Signatures of Heavy Neutrinos}
\label{sec:Collider}

The smoking-gun signature of a Majorana heavy neutrino at hadron colliders is the appearance of lepton number violating (LNV) same-sign (SS) dilepton events accompanied by jets. In contrast, a Dirac heavy neutrino preserves lepton number and contributes only to lepton number conserving (LNC) opposite-sign (OS) dilepton final states. Consequently, the relative abundance of SS and OS dilepton events provides a direct probe of the Majorana nature of the heavy neutrino state. To quantify possible departures from the pure Majorana or pure Dirac limits, we introduce an effective parameter $y \in [0,1]$, which characterizes the Majorana component of the heavy neutrino state. The limiting cases $y = 1$ and $y = 0$ correspond to purely Majorana and purely Dirac neutrinos, respectively. Under this parametrization, the observable ratio of SS to OS dilepton events can be expressed as

\begin{equation}
R_{\ell \ell} \equiv \frac{N_{\rm{SS}}}{N_{\rm{OS}}} = \frac{0.5 y \sigma_{\rm{Maj}}}{(1 -y) \sigma_{\rm{Dir}} + 0.5 y \sigma_{\rm{Maj}}},
\label{eq:Rll_mixed}
\end{equation}
where $\sigma_{\text{Maj}}$ and $\sigma_{\text{Dir}}$ denote the production cross sections for Majorana and Dirac-like heavy neutrino contributions, respectively. The factor of $1/2$ arises from the fact that a Majorana neutrino generates SS and OS dilepton final states with equal probability, whereas a Dirac neutrino contributes exclusively to the OS channel. Within this framework, the signal and background cross sections are modeled as weighted combinations of the pure Majorana and pure Dirac contributions,
\begin{align}
& \sigma^{\rm signal}_{\rm Mixed} = y \sigma^{\rm signal}_{\rm Maj} + ( 1 -y) \sigma^{\rm signal}_{\rm Dir}, \nonumber \\
& \sigma^{\rm Bkg}_{\rm Mixed} = y \sigma^{\rm Bkg}_{\rm Maj} + (1-y)\sigma^{\rm Bkg}_{\rm Dir}.
\end{align}
For a given integrated luminosity $\mathcal{L}$, the statistical sensitivity of the signal is estimated using the conventional significance measure
\begin{widetext}
\begin{equation}
\text{Significance}~(s) = \frac{\mathcal{L} \left[ (1 - y) \sigma^{\rm{signal}}_{\rm{Dir}} + y \sigma^{\rm{signal}}_{\rm{Maj}} \right] \epsilon^{\rm{signal}}}{\sqrt{ \mathcal{L} \left[ (1 - y) \sigma^{\rm{signal}}_{\rm{Dir}} + y \sigma^{\rm{signal}}_{\rm{Maj}} \right] \epsilon^{\rm{signal}} + \mathcal{L} \left[ (1 - y) \sigma^{\rm{Bkg}}_{\rm{Dir}} + y \sigma^{\rm{Bkg}}_{\rm{Maj}} \right] \epsilon^{\rm{Bkg}}}}~.
\label{eq:sig}
\end{equation}
\end{widetext}
Here $\epsilon^{\rm signal}$ and $\epsilon^{\rm Bkg}$ denote the corresponding signal and background efficiencies after event selection. In the numerical analysis, heavy neutrinos are produced through charged-current interactions at the 14 TeV LHC. We focus on the dominant decay channel $ N \rightarrow \ell^\pm jj$, leading to dilepton plus dijet final states. Both the LNV same-sign process,
\begin{equation}
    pp \rightarrow \ell^\pm N \rightarrow \ell^\pm \ell^\pm jj,
\end{equation}
and the LNC opposite-sign process,
\begin{equation}
 pp \rightarrow \ell^\pm N \rightarrow \ell^\pm \ell^\mp jj,
\end{equation}
are studied after parton showering, hadronization, and detector simulation. Event yields are evaluated for the current LHC integrated luminosity of $140~{\rm fb}^{-1}$ as well as the projected HL-LHC dataset of $3~{\rm ab}^{-1}$.

\subsection{Same-sign dilepton signal and background}
\label{subsec:SS}
To investigate the LNV same-sign dilepton signature of heavy neutrinos, we simulate the processes
\begin{equation}
    pp \to N \ell^\pm,~ N\to \ell^\pm W^\mp,~ W^\mp \to jj,
\end{equation}
where $\ell = e, \mu$. These channels lead to the characteristic SS dilepton plus dijet final state which constitutes a direct probe of the Majorana nature of heavy neutrinos. The dominant SM backgrounds arise from electroweak vector-boson scattering and top-quark pair production. In particular, same-sign $W^\pm W^\pm jj$ production represents an irreducible background with a production cross-section of $\sigma_{W^\pm W^\pm} = 119.26$ fb at $\sqrt{s} = 14$ TeV including NLO QCD and electroweak corrections~\cite{Biedermann:2017bss}. Another important contribution originates from $t\bar t$ production. Although this process predominantly yields OS dilepton final states, SS dilepton events can arise through charge misidentification and detector effects. To suppress this background, a $b$-jet veto is imposed assuming a $b$-tagging efficiency of $70 \%$ and a light-jet mistag probability of $1.5 \%$~\cite{CMS:2012feb}. Charge misidentification can convert OS dilepton events into apparent SS events. The corresponding effective SS cross section can be estimated as $\sigma_{\rm SS} \approx \sigma_{\rm OS}\times 2\,\epsilon_\ell $ where $\epsilon_\ell$ denotes the charge-misidentification probability. Throughout our analysis we adopt $\epsilon_e=10^{-3}$ and $\epsilon_\mu=10^{-5}$ for electrons and muons, respectively~\cite{CMS:2018ubm, ATLAS:2018ceg}. The $W^\pm W^\pm jj$ and charge-misidentified $t\bar t$ processes constitute the dominant backgrounds considered in this work.

To enhance the signal sensitivity, we adopt the following selection criteria:
\begin{itemize}
\item[(i)] Leading leptons $\ell^\pm$ have the transverse momentum $p^\ell_T > 20$ GeV, trailing lepton have the transverse momentum $p^\ell_T > 10$ GeV and leptons have the pseudo-rapidity range $|\eta^\ell| < 2.5$.
\item[(ii)] Hard jets having at least $p_T^j > 20 \,\mathrm{GeV}$, $|\eta^j| < 2.5$ and $\Delta R_{jj}=0.4$
\item[(iii)]The lepton–lepton separation: $\Delta R_{\ell \ell} > 0.4$ and the lepton-jet
separation: $\Delta R_{lj} > 0.4$.
\end{itemize}
In addition to the above criterions, we apply the following analysis specific selections:
\begin{itemize}
\item[(iv)] Only events with reconstructed two leptons having same-sign have been selected for the analysis.
\item[(v)] A mass window cut is imposed on the reconstructed invariant mass of the heavy neutrino candidate, $ \bigl| M_{\text{inv}} - M_N \bigr| < 0.2\, M_N $, where the chosen $(20\%)$ window accounts for the expected detector resolution at the HL-LHC, ensuring high signal efficiency while maintaining effective background rejection.
\item[(vi)] Absence of neutrinos in our signal, require us to consider the missing energy $E_T^{\rm{miss}} < 35$ GeV. It can control background events with large missing momenta.
\end{itemize}
The signal $(\epsilon^{\rm signal})$ and background $(\epsilon^{\rm Bkg})$ efficiencies are extracted from the ratio of the number of events surviving the selection cuts to the corresponding pre-selection yields. The resulting cut-flow efficiencies for representative heavy-neutrino benchmark masses are summarized in Tab.~\ref{tab:SScutflow}.
\begin{table*}[htb!]
\centering
\scriptsize
\resizebox{\textwidth}{!}{
\begin{tabular}{|c||c|c|c||c|c|c|}
\hline
\multirow{2}{*}{Cut}
& \multicolumn{3}{c||}{Signal for $M_N$}
& \multicolumn{3}{c|}{Background: $WWjj$}\\
\cline{2-7}
& 70 GeV & 120 GeV & 160 GeV
& 70 GeV & 120 GeV & 160 GeV \\
\hline
Pre-selection & 3073 + 3534 & 15803 + 19760 & 18409 + 23379 & 30144 + 38361 & 30144 + 38361 & 30144 + 38361 \\
+ (iv) & $[100 \%]$ & $[100 \%]$ & $[100 \%]$ & $[100 \%]$ &  $[100 \%]$ & $[100 \%]$ \\
\hline
Invariant Mass & 1843 + 2106 & 5987 + 7560 & 8740 + 11266 & 23 + 23 & 619 + 778 & 1815 + 2241  \\
& $[59.97 \% + 59.59 \%]$ & $[37.89 \% + 38.26 \%]$ & $[47.48 \% + 48.19 \%]$ & $[0.08 \% + 0.06 \%]$ &  $[2.05 \% + 2.03 \%]$ & $[6.02 \% + 5.84 \%$] \\
\hline
$E_T^{\rm{miss}} < 35$ GeV & 1843 + 2106  &5987 + 7560  &8739 + 11265 & 1+ 1 & 101 + 141 & 362 + 473 \\
& $[59.97 \% + 59.59 \%]$ & $[37.89 \% + 38.26 \%]$ & $[47.47 \% + 48.18 \%]$ & $[0.003 \% + 0.003 \%]$ &  $[0.34 \% + 0.37 \%]$ & $[1.20 \% + 1.23 \%$] \\
\hline
\end{tabular}
}
\caption{Cut-flow table for the same-sign dilepton signal and the dominant $WWjj$ background after successive event-selection requirements for benchmark heavy neutrino masses $M_N = 70$, 120, and 160 GeV. In each entry, the first (second) number corresponds to the $e^\pm e^\pm$ ($\mu^\pm \mu^\pm$) final state. The percentages shown in square brackets denote the efficiencies with respect to the pre-selection stage. Event yields are presented for an integrated luminosity of 140 fb$^{-1}$ and 3 ab$^{-1}$ (HL-LHC) at a proton-proton collider with center of mass energy $\sqrt{s}= 14$ TeV.}
\label{tab:SScutflow}
\end{table*}

\subsection{Opposite-sign dilepton signal and background}
\label{subsec:OS}

To investigate the LNC opposite-sign dilepton signature of heavy neutrinos, we simulate the processes
\begin{equation}
pp \to N \ell^\pm,~ N\to \ell^\mp W^\pm,~ W^\pm \to jj,
\end{equation}
where $\ell=e,\mu$. These channels give rise to the characteristic OS dilepton plus dijet final state. Unlike the SS channel, the OS signature is not a unique manifestation of the Majorana nature of heavy neutrinos and generally receives larger SM contributions. However, it plays an important role in determining the observable $R_{\ell\ell}$ through the relative abundance of same-sign and OS dilepton events.

In case of OS dilepton signature with two jets we study gauge boson decays to mimic the signal process. The dominant SM backgrounds arise from $t \bar{t}$, $Z+$ jets, and diboson production. Leptonic decays from $t\bar{t}$ will give a substantial contribution in the SM background, whose production cross section at $\sqrt{s} = 14$ TeV is approximately $\sigma_{t \bar{t}} = 835.61$ pb at N$^3$LO accuracy~\cite{Muselli:2015kba}. To suppress this background, a $b$-jet veto is imposed assuming a $b$-tagging efficiency of $70\%$ together with a light-jet mistag probability of $1.5\%$ \cite{CMS:2012feb}, similar to OS signal. A substantial background also arises from $Z+$jets production, where the neutral gauge boson decays into an OS dilepton pair. This contribution is effectively reduced by vetoing events whose dilepton invariant mass lies close to the $Z$ boson mass. In particular, we reject events satisfying $|M_{\ell \ell}-M_Z|\leq 5$ GeV. In addition, OS dilepton events can originate from diboson production. The corresponding production cross section for the process $ pp \to W^+W^-$ at NLO QCD accuracy is $\sigma_{W^+W^-} = 112.64$ pb~\cite{Campbell:2011bn}. These three processes constitute the dominant backgrounds considered in the present analysis.

To facilitate a direct comparison between the SS and OS channels, we employ the same selection criteria and isolation requirements described in Subsec.~\ref{subsec:SS}. In particular, identical lepton and jet transverse-momentum thresholds, pseudorapidity requirements, and angular-separation cuts are imposed. The OS analysis differs from the SS case only through one following analysis specific requirement:
\begin{itemize}
\item[(iv)] Only events with reconstructed two leptons having opposite-sign have been selected for the analysis.
\end{itemize}
We adopt the other two analysis specific selections similar to SS as described in Subsec.~\ref{subsec:SS}. The resulting cut-flow efficiencies of signal and background for representative benchmark masses are summarized in Tab.~\ref{tab:OScutflow}.
\begin{table*}[t]
\centering
\scriptsize
\resizebox{\textwidth}{!}{
\begin{tabular}{|c||c|c|c||c|c|c||c|c|c||c|c|c|}
\hline
\multirow{2}{*}{Cut}
& \multicolumn{3}{c||}{Signal for $M_N$}
& \multicolumn{3}{c||}{Background 1: $t\bar t$}
& \multicolumn{3}{c||}{Background 2: $WWjj$ }
& \multicolumn{3}{c|}{Background 3: $Z+\mathrm{jets}$}
\\
\cline{2-13}
& 70 & 120 & 160
& 70 & 120 & 160
& 70 & 120 & 160
& 70 & 120 & 160
\\
\hline

Pre-selection
& 2670+3171
& 14920+18928
& 16423+20958
& 40393+49894
& 40393+49894
& 40393+49894
& 34021+42600
& 34021+42600
& 34021+42600
& 28084+35609
& 28084+35609
& 28084+35609
\\
+(iv)
& $[100 \%]$
& $[100 \%]$
& $[100 \%]$
& $[100 \%]$
& $[100 \%]$
& $[100 \%]$
& $[100 \%]$
& $[100 \%]$
& $[100 \%]$
& $[100 \%]$
& $[100 \%]$
& $[100 \%]$
\\
\hline

Invariant Mass
& 1652+1912
& 5989+7478
& 8655+11132
& 49+59
& 1921+2378
& 7707+9542
& 153+154
& 2086+2558
& 4951+5738
& 2+3
& 2159+2694
& 5606+7085
\\
& $[61.90 \% + 60.30 \%]$
& $[40.10\% + 39.51 \%]$
& $[52.70 \% + 53.12 \%]$
& $[0.12 \% + 0.12 \%]$
& $[4.73 \% + 4.76 \%]$
& $[19 \% + 19.12 \%]$
& $[0.45 \% + 0.36 \%]$
& $[6.13 \% + 6.01\%]$
& $[13 \% + 13.46 \%]$
& $[0.01 \% + 0.01 \%]$
& $[7.69 \% + 7.57 \%]$
& $[19.96 \% + 19.90 \%]$
\\
\hline

$E_T^{\rm miss}<35~\mathrm{GeV}$
& 1652+1912
& 5989+7478
& 8655+11128
& 4+3
& 145+192
& 982+1260
& 14+18
& 502+590
& 1137+1439
& 2+3
& 2159+2694
& 5606+7085
\\
& $[61.90\% + 60.30 \%]$
& $[40.10 \% + 39.51 \%]$
& $[52.70 \% + 53.10 \%]$
& $[0.01 \% + 0.01 \%]$
& $[0.36 \% + 0.39 \%]$
& $[2.43 \% + 2.53 \%]$
& $[0.04 \% + 0.04 \%]$
& $[1.48 \% + 1.39\%]$
& $[3.34 \% + 3.38 \%]$
& $[0.01 \% + 0.01 \%]$
& $[7.69 \% + 7.57 \%]$
& $[19.96 \% + 19.90 \%]$
\\
\hline
\end{tabular}
}
\caption{Cut-flow table for the opposite-sign dilepton signal and the dominant Standard Model backgrounds after successive event-selection requirements for benchmark heavy-neutrino masses $M_N=70$, 120, and 160 GeV. The background processes considered are $WWjj$, $t\bar t$, and $Z+\mathrm{jets}$. The percentages in square brackets denote efficiencies relative to the pre-selection stage. Event yields are shown for an integrated luminosity of 140 fb$^{-1}$ and 3 ab$^{-1}$ at the HL-LHC with $\sqrt{s}=14$ TeV.}
\label{tab:OScutflow}
\end{table*}

\subsection{Numerical Results}
\label{subsec:numerical}
The statistical significance $(s)$ of the observed signal events $(S)$ over the total SM background events $(B)$ has been calculated using $\frac{S}{\sqrt{S+B}} = 2$. Throughout this work, we determine the parameter space corresponding to a $2 \sigma$ sensitivity. 
\begin{figure*}[htb!]
    \centering
\includegraphics[width=0.45\linewidth]{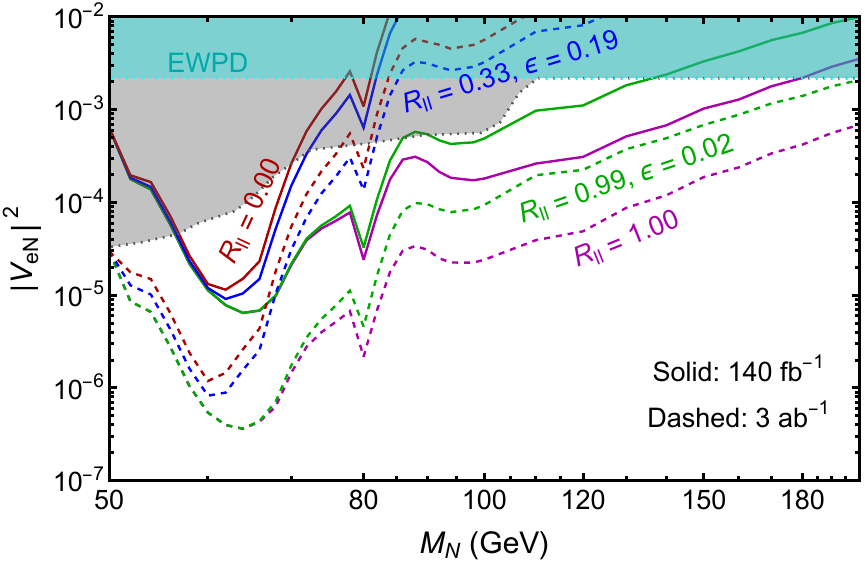} \qquad
\includegraphics[width=0.45\linewidth]{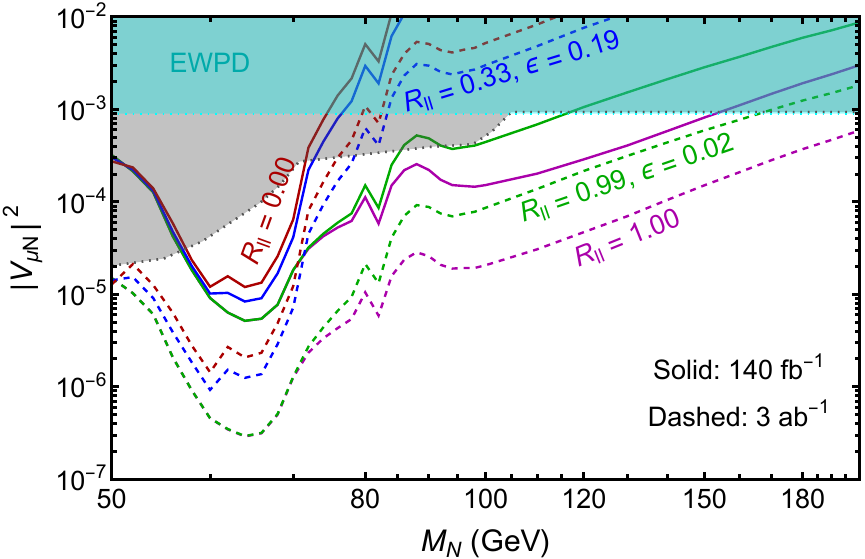} \qquad
\includegraphics[width=0.45\linewidth]{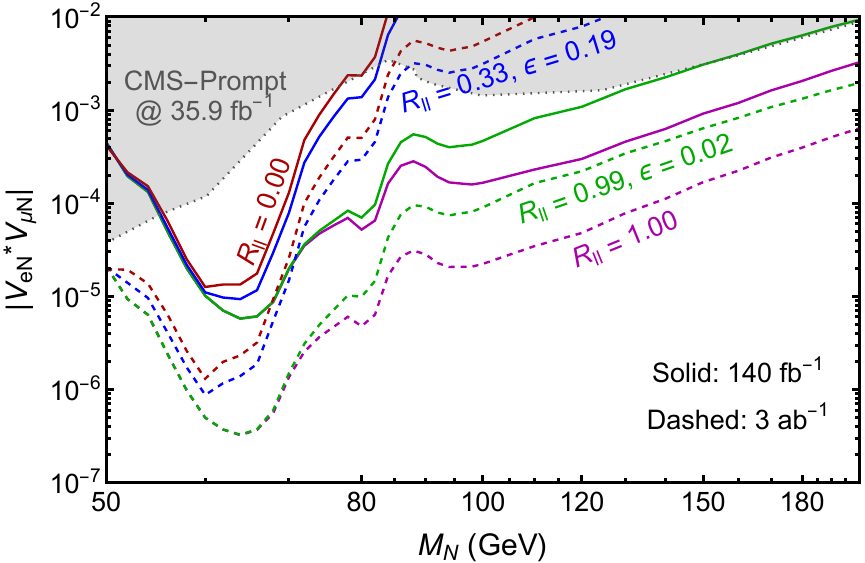} 
\caption{
Projected 95\% CL exclusion sensitivities to $|V_{eN}|^2$ (upper left), $|V_{\mu N}|^2$ (upper right), and $|V_{eN}V_{\mu N}^{\ast}|$ (lower) as functions of the heavy-neutrino mass $M_N$. The colored curves correspond to benchmark values of the dilepton ratio $(R_{\ell\ell}$: red $(0.00)$, blue $(0.33)$, green $(0.99)$, and magenta $(1.00)$. Solid and dashed curves represent projected sensitivities for integrated luminosities of $140~\mathrm{fb}^{-1}$ and $3~\mathrm{ab}^{-1}$, respectively. The optimal sensitivity is obtained for $M_N \simeq 60-80 ~\mathrm{GeV}$, where the HL-LHC can probe mixing strengths down to the $10^{-6}-10^{-7}$, improving upon existing direct bounds by up to two orders of magnitude. The shaded regions denote current exclusion limits from electroweak precision data (EWPD), and other collider searches. The values of the CP asymmetry has been derived using the combined central value of the Dirac CP phase $\delta = -1.08 \pi$ from T2K and NO$\nu$A~\cite{T2K:2025wet} experiments.}
\label{fig:sensitivity}
\end{figure*}
To derive the sensitivity bounds shown in Fig.~\ref{fig:sensitivity}, we parameterize the ratio of same-sign to total dilepton events by Eq.~\eqref{eq:Rll_mixed}. For a fixed value of $R_{\ell\ell}$, one can obtain the parameter $y$. The signal and background efficiencies can be calculated from Tabs.~\ref{tab:SScutflow}~and~\ref{tab:OScutflow}. Applying the signal and background efficiencies we estimate 2$\sigma$ limits on the mixing as a function of the mass of the heavy neutrinos following Eq.~\ref{eq:sig} for $e^\pm e^\pm$+2-jet, $\mu^\pm \mu^\pm$+2-jet and $e^\pm \mu^\pm$+2-jet which are shown in Fig.~\ref{fig:sensitivity}. The corresponding limits are $|V_{eN}|^2$, $|V_{\mu N}|^2$ and $|V_{eN}^{*}V_{\mu N}|$, respectively. Solid colored curves represent the expected sensitivities for the current LHC dataset of $140~\rm{fb}^{-1}$, whereas dashed colored curves show the projected reach at the HL-LHC with $3~\rm{ab}^{-1}$. 

The shaded regions include the already existing bounds from heavy neutrino searches at the LHC~\cite{ATLAS:2015gtp, CMS:2015qur, CMS:2018jxx, ATLAS:2019kpx, CMS:2018iaf} from the CMS and ATLAS experiments at $\sqrt{s}=8$ TeV and 13 TeV with different integrated luminosities for different leptonic flavors. These shaded region for $|V_{eN}|^2$ (top left) contains bounds obtained from the LEP experiment from~\cite{L3:2001zfe} in Fig.~\ref{fig:sensitivity}. The shaded regions of the top panel of Fig.~\ref{fig:sensitivity} contains bounds on the light-heavy mixing of the neutrinos for electron and muon flavors from the DELPHI~\cite{DELPHI:1996qcc} experiment, respectively. In the lower panel of Fig.~\ref{fig:sensitivity} we show the exclusion limits for $|V_{eN}^\ast V_{\mu N}|$ mixing in gray between the light-heavy neutrinos obtained from the $\sqrt{s}=13$ TeV LHC~\cite{CMS:2015qur, CMS:2024bni} at 35.9 fb$^{-1}$ luminosity. Exclusion limits from the electroweak precision data (EWPD) in the upper panel of Fig.~\ref{fig:sensitivity} are taken from~\cite{delAguila:2008pw, deBlas:2013gla, Akhmedov:2013hec}. 

The projected exclusion sensitivities shown in Fig.~\ref{fig:sensitivity} exhibit their strongest reach in the mass interval $M_N \simeq 60-80~\rm{GeV}$, where the interplay between relatively large electroweak production rates and favorable decay kinematics maximizes the signal acceptance. In this region, the projected sensitivity for the current LHC dataset of $140~\rm{fb}^{-1}$ reaches approximately $|V_{\ell N}|^2 \sim 10^{-5}$, while the HL-LHC projection with an integrated luminosity of $3~\rm{ab}^{-1}$ extends the reach to $ \mathcal{O}(10^{-6})$. In particular, around $M_N \simeq 70~\rm{GeV}$ the HL-LHC is expected to probe $|V_{eN}|^2$ and $|V_{\mu N}|^2$ down to approximately $(5\times10^{-7}-10^{-6})$, depending on the benchmark values of $R_{\ell\ell}$. This corresponds to an improvement of approximately one to two orders of magnitude relative to the existing direct-search constraints shown by the shaded regions. The structures visible around $M_N \simeq 80-90~\rm{GeV}$ originate from electroweak gauge-boson thresholds. As the heavy RHN approaches the on-shell $W$- and $Z$-boson production regions, both the available phase space and the decay branching fractions undergo significant changes, leading to the characteristic features observed in the sensitivity curves. Beyond this threshold region, the sensitivity gradually deteriorates with increasing mass owing to the rapidly decreasing heavy-neutrino production cross section. However, the HL-LHC retains substantial discovery potential throughout the mass range considered and significantly improves upon current bounds even for $M_N \gtrsim 100~\rm{GeV}$.

The different colored curves correspond to benchmark scenarios characterized by distinct values of the dilepton ratio $R_{\ell\ell}$. As discussed previously, the relation between the dilepton ratio and the CP-asymmetry parameter, given in Eq.~\eqref{eq:RllEpsilon}, is applicable only in the mixed-state regime. Moreover, the CP asymmetry is highly sensitive to the value of the Dirac CP-violating phase. Therefore, for a benchmark scenario, we have mainly focused on the combined central value of the Dirac CP phase $\delta = -1.08 \pi$ from T2K and NO$\nu$A~\cite{T2K:2025wet} experiments. Consequently, the corresponding CP asymmetries are shown only for the physically relevant benchmark points $R_{\ell\ell}=0.33$ and $R_{\ell\ell}=0.99$, for which we obtain $\epsilon \simeq 0.19$ and $\epsilon \simeq 0.02$, respectively~\footnote{The correspondence between $R_{\ell\ell}$ and $\epsilon$ generally depends on the $|V_{\ell N}|^2$. Nevertheless, for the benchmark point with $\delta=-1.08\pi$, we find that this dependence is weak in our interested range $10^{-6} \lesssim |V_{\ell N}|^2 \lesssim 10^{-2}$.}. We find that, with the benchmark scenario $R_{\ell\ell}=1$ yielding the strongest exclusion reach, followed by $R_{\ell\ell}=0.99$ and $R_{\ell\ell}=0.33$, while the purely Dirac limit $R_{\ell\ell}=0$ generally leads to the weakest constraints. Around $M_N \simeq 70~\rm{GeV}$, the difference between the strongest and weakest projected sensitivities approaches nearly one order of magnitude. This behavior shows that the collider sensitivity is closely linked to the underlying heavy-neutrino dynamics. Since $R_{\ell\ell}$ determines the relative contributions of LNC and LNV processes, different values of $R_{\ell\ell}$ can lead to noticeably different exclusion limits on the light-heavy mixing parameters. Different values of $R_{\ell\ell}$ therefore correspond to distinct regions of the $(R_{\ell\ell}, \epsilon)$ parameter space and lead to different collider signatures. The benchmark curves shown in Fig.~\ref{fig:sensitivity} may consequently be interpreted not only as collider exclusion limits but also as indirect probes of the CP-violating dynamics responsible for generating the baryon asymmetry of the Universe.

Moreover, a comparison of the electron and muon channels shows the importance of flavor-dependent effects. Even though the upper-left and upper-right panels follow similar overall trends across the benchmark scenarios of $R_{\ell\ell}$, small differences appear because of different detector efficiencies, and background processes for electrons and muons. In both channels, the best sensitivity is found in the low-mass region, where the projected limits reach about the $10^{-6}$ level at the HL-LHC. The close agreement between the electron and muon results for $|V_{eN}|^2$ and $|V_{\mu N}|^2$ suggests that future experiments will be similarly sensitive to heavy neutrinos that couple mainly to either electrons or muons, consistent with different values of $R_{\ell\ell}$. Unlike the single-flavor cases, the mixed-flavor quantity $|V_{eN}V_{\mu N}^{*}|$, shown in the lower panel of Fig.~\ref{fig:sensitivity} depends on both electron and muon couplings at the same time, so it directly reflects the flavor structure of the neutrino mass model. Around $M_N \simeq 70~\rm{GeV}$, the HL-LHC can probe values down to the few-$10^{-7}$ level, making it sensitive to scenarios where heavy neutrinos couple to more than one lepton flavor.

Overall, the projected sensitivities presented in Fig.~\ref{fig:sensitivity} establish a direct and experimentally testable connection between collider searches for heavy neutral leptons, the flavor structure of light-heavy neutrino mixing, low-energy neutrino oscillation measurements, and the CP-violating mechanism responsible for the generation of the baryon asymmetry of the Universe. The HL-LHC will therefore not only extend the search reach for heavy neutral leptons by up to two orders of magnitude beyond current limits, but also provide a unique opportunity to explore the flavor and CP properties of the underlying neutrino sector.

%%%%%%%%%%%%%%%%%%%%%%%%%%%%%%%%%%%%%%%%%%
\section{Conclusions}
\label{sec:Conclusion}
%%%%%%%%%%%%%%%%%%%%%%%%%%%%%%%%%%%%%%%%%%
In this paper, we study the type-I seesaw scenario in which the SM is extended by SM-singlet right-handed neutrinos (RHNs). In this framework, light neutrino masses arise naturally via the seesaw mechanism after electroweak symmetry breaking, while the heavy neutrino states become Majorana particles. The mixing between light and heavy neutrino mass eigenstates induces effective charged- and neutral-current interactions between the heavy neutrinos and SM gauge bosons.

This light-heavy mixing enables the production of heavy neutrinos at the 14 TeV LHC in association with a charged lepton, followed by their decay into dilepton plus dijet final states, yielding characteristic same-sign (SS) and opposite-sign (OS) signatures. The SS channel typically suffers from smaller SM backgrounds due to its lepton number violating nature, whereas the OS channel receives larger backgrounds from SM processes. We therefore perform a cut-based detector-level analysis of both channels and their dominant backgrounds as a function of the ratio $R_{\ell\ell}$ of SS to OS dilepton events. Using this framework, we derive constraints on the light-heavy neutrino mixing as a function of the heavy neutrino mass. We find that, in the most favorable case with ($R_{\ell\ell}\sim 1$), the sensitivity at 140~$\rm{fb}^{-1}$ can surpass current LHC limits by about one order of magnitude for $M_N \simeq 60-80~\rm{GeV}$, with a further improvement of nearly another order of magnitude expected at the HL-LHC with $3000~\rm{fb}^{-1}$. For smaller values of $R_{\ell\ell}$, the sensitivity can drop by up to two orders of magnitude.

As emphasized in Ref.~\cite{Drewes:2019byd}, for $M_N \gtrsim 100~\rm{GeV}$, radiative stability of the light neutrino masses requires $R_{\ell\ell} < 1/3$. In this region, our projected sensitivity, even at the HL-LHC, does not reach the level of current experimental bounds. By contrast, for $50~\rm{GeV} \lesssim M_N \lesssim 100~\rm{GeV}$, no such requirement on $R_{\ell\ell}$ applies. In this mass range, we show that the sensitivity depends strongly on $R_{\ell\ell}$, and hence also on the CP asymmetry $\epsilon$, for representative benchmark choices in the Casas--Ibarra parametrization. This provides additional motivation for future experimental efforts to determine $R_{\ell\ell}$, for instance through searches for heavy-neutrino oscillations at colliders.

%%%%%%%%%%%%%%%%%%%%%%%%%%%%%%%%%%%%%%
\vspace{10pt}
\textbf{Acknowledgements} \\
%%%%%%%%%%%%%%%%%%%%%%%%%%%%%%%%%%%%%%%%%%%%
We thank Chayan Majumdar for useful discussions. W. L. is supported by National Natural Science foundation of China (Grant No. 12205153).

%%%%%%%%%%%%%%%%%%%%%%%%%%%%%%%%%%%%%%%%%%%%%%%
\appendix
\section{Heavy Neutrino Decays}
\label{app:Ndecays}
%%%%%%%%%%%%%%%%%%%%%%%%%%%%%%%%%%%%%%%%%%%%%%%
The two-body partial decay widths of heavy neutrinos considering $M_N > M_{W,Z,h}$ are given by
\begin{align}
\Gamma(N_i \to \ell_\alpha W) & = \frac{g^2 |V_{\alpha i}|^2}{64\pi}
\frac{(M_{N_i}^2 - M_W^2)^2 (M_{N_i}^2 + 2 M_W^2)} {M_{N_i}^3 M_W^2},
\\
\Gamma(N_i \to \nu_\alpha Z) & =
\frac{g^2 |V_{\alpha i}|^2}{128\pi c_w^2} \frac{(M_{N_i}^2 - M_Z^2)^2 (M_{N_i}^2 + 2 M_Z^2)} {M_{N_i}^3 M_Z^2} ,
\\
\Gamma(N_i \to \nu_\alpha h) & = \frac{|V_{\alpha i} |^2}{32\pi} \frac{(M_{N_i}^2 - M_h^2)^2}{M_{N_i} v^2},
\end{align}
In the limit $M_{N_i} \gg M_{W,Z,h}$, the partial widths approach the asymptotic ratio
\begin{equation}
\resizebox{\columnwidth}{!}{$
\Gamma(N_i \to \ell_\alpha W) :
\Gamma(N_i \to \nu Z) :
\Gamma(N_i \to \nu h)
\simeq 2 : 1 : 1 ,
$}
\end{equation}
reflecting the enhancement of the charged-current channel associated with the two transverse degrees of freedom of the $W^\pm$ boson. For $M_{N_i} < M_W$, two-body decays into on-shell gauge bosons are kinematically forbidden. In this regime, heavy neutrinos decay predominantly via three-body~\cite{Hagedorn:1963hdh, Nachtmann:1990ta, Byckling:1971vca, Chun:2019nwi, A:2025ygb} processes mediated by off-shell electroweak gauge bosons. Considering virtual $W$ boson decays into leptons for $M_N < M_W$, partial decay width is given by
\begin{equation}
\resizebox{\columnwidth}{!}{$
\Gamma(N \to \ell_1^- \ell_2^+ \nu_{\ell_2}) = \frac{G_F^2 M_N^5 |V_{\ell_1 N}|^2 }{16\pi^3}  (1-\delta_{\ell_1 \ell_2}) I_1 (y_{\nu_{\ell_2}}, y_{\ell_1}, y_{\ell_2}), 
\label{eq:nul1l2}
$}
\end{equation}
with $G_F = 1.166\times 10^{-5}$ GeV$^{-2}$ being the Fermi constant. Here $\delta_{\ell_1 \ell_2}$ is the Kronecker delta and $y_i \equiv m_i / M_N$ where $m_i$ is the mass of the corresponding lepton. In case of a virtual $Z$ boson decays into a pair of charged leptons, the partial decay width is given by
\begin{align}
\Gamma(N\to \nu_{\ell_1} \ell_{2}^{-} \ell_{2}^{+}) = &
\frac{G_F^2 M_N^5 |V_{\ell_1 N}|^2}{8\pi^3}  \Big[2(g_L^\ell g_R^\ell + g_R^\ell\delta_{\ell_1 \ell_2}) \nonumber \\
& 
I_2 (y_{\nu_{\ell_1}}, y_{\ell_2}, y_{\ell_2}) + \bigg[ (g_L^\ell)^2+(g_R^\ell)^2 \nonumber \\
&
+(1 + 2 g_L^\ell) \delta_{\ell_1 \ell_2} \bigg]  I_1(y_{\nu_{\ell_1}}, y_{\ell_2}, y_{\ell_2})\Big], 
\label{eq:nuiljlj}
\end{align}
where $g_L^\ell = - 1/2 + \sin^2\theta_{\rm W}$ and $g_R^\ell = \sin^2\theta_{\rm W}$, respectively. Note that in the above case we also add the process $N \to \bar{\nu}_{\ell_1} \ell_2^-\ell_2^+$. Heavy neutrinos can decay into three light neutrinos through a virtual $Z$ boson. The corresponding partial decay width can be given by
\begin{equation}
\sum_{\ell_2=e,\mu,\tau}^{} \Gamma(N\to \nu_{\ell_1}\nu_{\ell_2}\bar{\nu}_{\ell_2})=|V_{\ell_1 N}|^2\frac{G_F^2}{96\pi^3}M_N^5~.
\label{eq:3nu}
\end{equation}
adding the process $N \to \bar{\nu}_{\ell_1} \nu_{\ell_2} \bar{\nu}_{\ell_2}$ in Eq.~\eqref{eq:3nu} ignoring the neutrino mass due to smallness. Partial decay widths of heavy neutrinos into charged and neutral pseudoscalar mesons $(P^{+,0})$ can be given as
\begin{align}
\Gamma(N \to \ell^-_{1} P^+) & = |V_{\ell_1 N}|^2 \frac{G_F^2}{16\pi} M_N^3 f_P^2 |V_P|^2 F_P(y_{\ell}, y_P),
\label{eq:l1p} \\
\Gamma(N\to \nu_{\ell_{1}} P^0) & = |V_{\ell_1 N}|^2 \frac{G_F^2}{2\pi}M_N^3f_P^2 \kappa_P^2 F_P(y_{\nu_\ell},y_P),
\end{align}
where $f_P$ represents decay constant of the pseudoscalar $P$ and $V_P$ is the corresponding elements of the Cabibbo–Kobayashi– Maskawa (CKM) matrix as listed in Tab.~\ref{tab:mesdec} for each meson.
\begin{table*}[t]
\centering
\begin{tabular}{|c|c|c|c||c|c|c|c|}
\hline
Psudoscalar(P)  & $m_P$ (MeV) & $f_P$(MeV) & $V_P$ & Vector(V)  & $m_V$ (MeV) & $f_V$(MeV) & $V_V$ \\ 
\hline
$\pi^\pm$ & 139.6 & 130.7 & $V_{ud}$ & $\rho^\pm$ & 775.8 & 220 & $V_{ud}$ \\
\hline
$K^\pm$ & 493.7 & 159.8 & $V_{us}$ & $K^{*\pm}$ & 891.66 & 217 & $V_{us}$ \\
\hline
$\eta$ & 547.8 & 164.7 & -- & $\omega$ & 782.59 & 195 & -- \\ 
\hline
$\pi^0$ & 135 & 130 & -- & $\rho^0$ & 776 & 220 & -- \\ 
\hline
$K^0$ & 497.6 & 159 & -- & $K^{*0}$ & 896.1 & 217  & -- \\ 
\hline
\end{tabular}
\caption{Masses and decay constants of pseudoscalar and vector mesons following~\cite{ParticleDataGroup:2010dbb, CLEO:2005jsh, MILC:2002lnl, Ivanov:2006ni, Feldmann:1999uf, Ebert:2006hj} used in the three body decays of the heavy neutrinos.}
\label{tab:mesdec}
\end{table*}
Due to the NC coupling, $\kappa_P$ of the pseudoscalar meson is defined as
\begin{align}
\kappa_{\pi^{0}} = -\frac{1}{2\sqrt{2}}, \kappa_{\eta}=-\frac{1}{2\sqrt{6}}, \rm{and}~ \kappa_{K^0} = \frac{1}{4}.
\end{align}
Partial decay widths of heavy neutrinos into charged and neutral vector mesons $(V^{+,0})$ are given by
\begin{align}
\Gamma(N\to \ell^-_1 V^+) & = |V_{\ell_1 N}|^2 \frac{G_F^2}{16\pi} M_N^3 f_V^2 |V_V|^2 F_V(y_{\ell}, y_V), 
\label{eq:l1v} \\
\Gamma(N\to \nu_{\ell_1}V^0) & = |V_{\ell_1 N}|^2 \frac{G_F^2}{2 \pi} M_N^3 f_V^2 \kappa_V^2 F_V( y_{\nu_\ell}, y_V),
\label{eq:nuv}
\end{align}
where $f_V$ represents the decay constant of the vector meson $V$. Here $\kappa_V$ is the corresponding NC coupling of vector meson and they are defined as
\begin{align}
& \kappa_{\rho^{0}} = \frac{1}{\sqrt{2}} \left( \frac{1}{2} -\rm{sin}^{2} \theta_{W}  \right),~ 
\kappa_{\omega} = - \frac{1}{3 \sqrt{2}} \rm{sin}^{2} \theta_{W}, \nonumber \\
& \rm{and}~ \kappa_{K^{*0}} = \frac{1}{2} \left( \frac{2}{3} \rm{sin}^{2} \theta_{W} - \frac{1}{2} \right).
\end{align}

Heavy neutrinos decay into opposite sign different flavored quarks through virtual $W$ boson and the corresponding partial decay width is given by 
\begin{equation}
\Gamma(N \to \ell_1^- q \bar{q}^\prime ) = N_C |V_{\ell_1 N}|^2 |V_{q\bar{q}^\prime}|^2 \frac{G_F^2}{16\pi^3} M_N^5 I_1( y_{\nu_{\ell_1}}, y_q, y_{q'}),
\label{eq:l1ud}
\end{equation}
where $V_{q\bar{q}^\prime}$ is the corresponding element of the CKM matrix as listed in Tab.~\ref{tab:mesdec}.  The Similar interaction could be possible through a virtual $Z$ boson exchange, and the corresponding partial decay width is given by
\begin{align}
 \Gamma(N \to \nu_{\ell_1} q \bar{q}) & = N_C |V_{\ell_1 N}|^2 \frac{G_F^2}{8\pi^3} M_N^5 \bigg[2 g_L^q g_R^q I_2(y_{\nu_{\ell}}, y_q, y_q) \nonumber \\
 & +((g_L^q)^2+(g_R^q)^2) I_1(y_{\nu_{\ell}}, y_q, y_q)\bigg],
\label{eq:nuqq}
\end{align}
where $ g_L^u = 1/2 - (2/3) \sin^2 \theta_{\rm W}$, $g_R^u = - (2/3) \sin^2 \theta_{\rm W}$, $g_L^d = -1/2 + (1/3) \sin^2 \theta_{\rm W}$, and $g_R^d = (1/3) \sin^2 \theta_{\rm W}$, respectively. In the above formulae of the partial decay widths of RHNs, the kinematical functions are give by
\begin{align}
I_1(x,y,z) & = 
\int_{(x+y)^2}^{(1-z)^2}\frac{ds}{s}(s-x^2-y^2)(1+z^2-s) \nonumber \\
&~~~~~~~~~~~~~~~~~~~ 
\lambda(s,x^2,y^2) \lambda(1,s,z^2), \\
I_2(x,y,z) & = y z \int_{(y+z)^2}^{(1-x)^2}  \frac{ds}{s}(1+x^2-s) \nonumber \\
& ~~~~~~~~~~~~~~~~~~~~~ 
\lambda(s,y^2,z^2) \lambda(1,s,x^2), \\
F_P(x,y) & = \lambda(1,x^2,y^2)[(1+x^2)(1+x^2-y^2)-4x^2], \\
F_V(x,y) &= \lambda(1,x^2,y^2) [(1-x^2)^2+(1+x^2)y^2-2y^4],
\end{align}
with the K\"{a}ll\'{e}n function
\begin{equation}
\lambda(x,y,z) = \sqrt{x^2+y^2+z^2-2xy-2yz-2zx}~.
\end{equation}

\bibliographystyle{utcaps_mod}
\bibliography{Reference}

\providecommand{\href}[2]{#2}\begingroup\raggedright\begin{thebibliography}{10}

\bibitem{ParticleDataGroup:2024cfk}
{\bfseries Particle Data Group} Collaboration, S.~Navas {\em et~al.}, ``{Review of particle physics},'' \href{http://dx.doi.org/10.1103/PhysRevD.110.030001}{{\em Phys. Rev. D} {\bfseries 110} no.~3, (2024) 030001}.

\bibitem{Yanagida:1980xy}
T.~Yanagida, ``{Horizontal Symmetry and Masses of Neutrinos},'' \href{http://dx.doi.org/10.1143/PTP.64.1103}{{\em Prog. Theor. Phys.} {\bfseries 64} (1980) 1103}.

\bibitem{Minkowski:1977sc}
P.~Minkowski, ``{$\mu \to e\gamma$ at a Rate of One Out of $10^{9}$ Muon Decays?},'' \href{http://dx.doi.org/10.1016/0370-2693(77)90435-X}{{\em Phys. Lett. B} {\bfseries 67} (1977) 421--428}.

\bibitem{Yanagida:1979as}
T.~Yanagida, ``{Horizontal gauge symmetry and masses of neutrinos},'' {\em Conf. Proc. C} {\bfseries 7902131} (1979) 95--99.

\bibitem{Gell-Mann:1979vob}
M.~Gell-Mann, P.~Ramond, and R.~Slansky, ``{Complex Spinors and Unified Theories},'' {\em Conf. Proc. C} {\bfseries 790927} (1979) 315--321, \href{http://arxiv.org/abs/1306.4669}{[{\ttfamily 1306.4669}]}.

\bibitem{Glashow:1979nm}
S.~L. Glashow, ``{The Future of Elementary Particle Physics},'' \href{http://dx.doi.org/10.1007/978-1-4684-7197-7_15}{{\em NATO Sci. Ser. B} {\bfseries 61} (1980) 687}.

\bibitem{Mohapatra:1979ia}
R.~N. Mohapatra and G.~Senjanovic, ``{Neutrino Mass and Spontaneous Parity Nonconservation},'' \href{http://dx.doi.org/10.1103/PhysRevLett.44.912}{{\em Phys. Rev. Lett.} {\bfseries 44} (1980) 912}.

\bibitem{Schechter:1980gr}
J.~Schechter and J.~W.~F. Valle, ``{Neutrino Masses in SU(2) x U(1) Theories},'' \href{http://dx.doi.org/10.1103/PhysRevD.22.2227}{{\em Phys. Rev. D} {\bfseries 22} (1980) 2227}.

\bibitem{Weinberg:1979sa}
S.~Weinberg, ``{Baryon and Lepton Nonconserving Processes},'' \href{http://dx.doi.org/10.1103/PhysRevLett.43.1566}{{\em Phys. Rev. Lett.} {\bfseries 43} (1979) 1566--1570}.

\bibitem{Casas:2001sr}
J.~A. Casas and A.~Ibarra, ``{Oscillating neutrinos and $\mu \to e, \gamma$},'' \href{http://dx.doi.org/10.1016/S0550-3213(01)00475-8}{{\em Nucl. Phys. B} {\bfseries 618} (2001) 171--204}, \href{http://arxiv.org/abs/hep-ph/0103065}{[{\ttfamily hep-ph/0103065}]}.

\bibitem{Antusch:2006vwa}
S.~Antusch, C.~Biggio, E.~Fernandez-Martinez, M.~B. Gavela, and J.~Lopez-Pavon, ``{Unitarity of the Leptonic Mixing Matrix},'' \href{http://dx.doi.org/10.1088/1126-6708/2006/10/084}{{\em JHEP} {\bfseries 10} (2006) 084}, \href{http://arxiv.org/abs/hep-ph/0607020}{[{\ttfamily hep-ph/0607020}]}.

\bibitem{Abada:2007ux}
A.~Abada, C.~Biggio, F.~Bonnet, M.~B. Gavela, and T.~Hambye, ``{Low energy effects of neutrino masses},'' \href{http://dx.doi.org/10.1088/1126-6708/2007/12/061}{{\em JHEP} {\bfseries 12} (2007) 061}, \href{http://arxiv.org/abs/0707.4058}{[{\ttfamily 0707.4058}]}.

\bibitem{MEG:2011naj}
{\bfseries MEG} Collaboration, J.~Adam {\em et~al.}, ``{New limit on the lepton-flavour violating decay $\mu^{+} \to e^{+} \gamma$},'' \href{http://dx.doi.org/10.1103/PhysRevLett.107.171801}{{\em Phys. Rev. Lett.} {\bfseries 107} (2011) 171801}, \href{http://arxiv.org/abs/1107.5547}{[{\ttfamily 1107.5547}]}.

\bibitem{BaBar:2009hkt}
{\bfseries BaBar} Collaboration, B.~Aubert {\em et~al.}, ``{Searches for Lepton Flavor Violation in the Decays $\tau^\pm \to e^\pm \gamma$ and $\tau^\pm \to \mu^\pm \gamma$},'' \href{http://dx.doi.org/10.1103/PhysRevLett.104.021802}{{\em Phys. Rev. Lett.} {\bfseries 104} (2010) 021802}, \href{http://arxiv.org/abs/0908.2381}{[{\ttfamily 0908.2381}]}.

\bibitem{SuperB:2010cqs}
{\bfseries SuperB} Collaboration, B.~O'Leary {\em et~al.}, ``{SuperB Progress Reports -- Physics},'' \href{http://arxiv.org/abs/1008.1541}{[{\ttfamily 1008.1541}]}.

\bibitem{WMAP:2010qai}
{\bfseries WMAP} Collaboration, E.~Komatsu {\em et~al.}, ``{Seven-Year Wilkinson Microwave Anisotropy Probe (WMAP) Observations: Cosmological Interpretation},'' \href{http://dx.doi.org/10.1088/0067-0049/192/2/18}{{\em Astrophys. J. Suppl.} {\bfseries 192} (2011) 18}, \href{http://arxiv.org/abs/1001.4538}{[{\ttfamily 1001.4538}]}.

\bibitem{Cohen:1993nk}
A.~G. Cohen, D.~B. Kaplan, and A.~E. Nelson, ``{Progress in electroweak baryogenesis},'' \href{http://dx.doi.org/10.1146/annurev.ns.43.120193.000331}{{\em Ann. Rev. Nucl. Part. Sci.} {\bfseries 43} (1993) 27--70}, \href{http://arxiv.org/abs/hep-ph/9302210}{[{\ttfamily hep-ph/9302210}]}.

\bibitem{Funakubo:1996dw}
K.~Funakubo, ``{CP violation and baryogenesis at the electroweak phase transition},'' \href{http://dx.doi.org/10.1143/PTP.96.475}{{\em Prog. Theor. Phys.} {\bfseries 96} (1996) 475--520}, \href{http://arxiv.org/abs/hep-ph/9608358}{[{\ttfamily hep-ph/9608358}]}.

\bibitem{Trodden:1998ym}
M.~Trodden, ``{Electroweak baryogenesis},'' \href{http://dx.doi.org/10.1103/RevModPhys.71.1463}{{\em Rev. Mod. Phys.} {\bfseries 71} (1999) 1463--1500}, \href{http://arxiv.org/abs/hep-ph/9803479}{[{\ttfamily hep-ph/9803479}]}.

\bibitem{CMS:2012qbp}
{\bfseries CMS} Collaboration, S.~Chatrchyan {\em et~al.}, ``{Observation of a New Boson at a Mass of 125 GeV with the CMS Experiment at the LHC},'' \href{http://dx.doi.org/10.1016/j.physletb.2012.08.021}{{\em Phys. Lett. B} {\bfseries 716} (2012) 30--61}, \href{http://arxiv.org/abs/1207.7235}{[{\ttfamily 1207.7235}]}.

\bibitem{ATLAS:2012yve}
{\bfseries ATLAS} Collaboration, G.~Aad {\em et~al.}, ``{Observation of a new particle in the search for the Standard Model Higgs boson with the ATLAS detector at the LHC},'' \href{http://dx.doi.org/10.1016/j.physletb.2012.08.020}{{\em Phys. Lett. B} {\bfseries 716} (2012) 1--29}, \href{http://arxiv.org/abs/1207.7214}{[{\ttfamily 1207.7214}]}.

\bibitem{Fukugita:1986hr}
M.~Fukugita and T.~Yanagida, ``{Baryogenesis Without Grand Unification},'' \href{http://dx.doi.org/10.1016/0370-2693(86)91126-3}{{\em Phys. Lett. B} {\bfseries 174} (1986) 45--47}.

\bibitem{Manton:1983nd}
N.~S. Manton, ``{Topology in the Weinberg-Salam Theory},'' \href{http://dx.doi.org/10.1103/PhysRevD.28.2019}{{\em Phys. Rev. D} {\bfseries 28} (1983) 2019}.

\bibitem{Klinkhamer:1984di}
F.~R. Klinkhamer and N.~S. Manton, ``{A Saddle Point Solution in the Weinberg-Salam Theory},'' \href{http://dx.doi.org/10.1103/PhysRevD.30.2212}{{\em Phys. Rev. D} {\bfseries 30} (1984) 2212}.

\bibitem{Khlebnikov:1988sr}
S.~Y. Khlebnikov and M.~E. Shaposhnikov, ``{The Statistical Theory of Anomalous Fermion Number Nonconservation},'' \href{http://dx.doi.org/10.1016/0550-3213(88)90133-2}{{\em Nucl. Phys. B} {\bfseries 308} (1988) 885--912}.

\bibitem{Pilaftsis:1997dr}
A.~Pilaftsis, ``{Resonant CP violation induced by particle mixing in transition amplitudes},'' \href{http://dx.doi.org/10.1016/S0550-3213(97)00469-0}{{\em Nucl. Phys. B} {\bfseries 504} (1997) 61--107}, \href{http://arxiv.org/abs/hep-ph/9702393}{[{\ttfamily hep-ph/9702393}]}.

\bibitem{Bray:2007ru}
S.~Bray, J.~S. Lee, and A.~Pilaftsis, ``{Resonant CP violation due to heavy neutrinos at the LHC},'' \href{http://dx.doi.org/10.1016/j.nuclphysb.2007.07.002}{{\em Nucl. Phys. B} {\bfseries 786} (2007) 95--118}, \href{http://arxiv.org/abs/hep-ph/0702294}{[{\ttfamily hep-ph/0702294}]}.

\bibitem{Deppisch:2015qwa}
F.~F. Deppisch, P.~S. Bhupal~Dev, and A.~Pilaftsis, ``{Neutrinos and Collider Physics},'' \href{http://dx.doi.org/10.1088/1367-2630/17/7/075019}{{\em New J. Phys.} {\bfseries 17} no.~7, (2015) 075019}, \href{http://arxiv.org/abs/1502.06541}{[{\ttfamily 1502.06541}]}.

\bibitem{Das:2015toa}
A.~Das and N.~Okada, ``{Improved bounds on the heavy neutrino productions at the LHC},'' \href{http://dx.doi.org/10.1103/PhysRevD.93.033003}{{\em Phys. Rev. D} {\bfseries 93} no.~3, (2016) 033003}, \href{http://arxiv.org/abs/1510.04790}{[{\ttfamily 1510.04790}]}.

\bibitem{Das:2016hof}
A.~Das, P.~Konar, and S.~Majhi, ``{Production of Heavy neutrino in next-to-leading order QCD at the LHC and beyond},'' \href{http://dx.doi.org/10.1007/JHEP06(2016)019}{{\em JHEP} {\bfseries 06} (2016) 019}, \href{http://arxiv.org/abs/1604.00608}{[{\ttfamily 1604.00608}]}.

\bibitem{Bolton:2019pcu}
P.~D. Bolton, F.~F. Deppisch, and P.~S. Bhupal~Dev, ``{Neutrinoless double beta decay versus other probes of heavy sterile neutrinos},'' \href{http://dx.doi.org/10.1007/JHEP03(2020)170}{{\em JHEP} {\bfseries 03} (2020) 170}, \href{http://arxiv.org/abs/1912.03058}{[{\ttfamily 1912.03058}]}.

\bibitem{Kersten:2007vk}
J.~Kersten and A.~Y. Smirnov, ``{Right-Handed Neutrinos at CERN LHC and the Mechanism of Neutrino Mass Generation},'' \href{http://dx.doi.org/10.1103/PhysRevD.76.073005}{{\em Phys. Rev. D} {\bfseries 76} (2007) 073005}, \href{http://arxiv.org/abs/0705.3221}{[{\ttfamily 0705.3221}]}.

\bibitem{Drewes:2019byd}
M.~Drewes, J.~Klari{\'c}, and P.~Klose, ``{On lepton number violation in heavy neutrino decays at colliders},'' \href{http://dx.doi.org/10.1007/JHEP11(2019)032}{{\em JHEP} {\bfseries 11} (2019) 032}, \href{http://arxiv.org/abs/1907.13034}{[{\ttfamily 1907.13034}]}.

\bibitem{Asaka:2005pn}
T.~Asaka and M.~Shaposhnikov, ``{The $\nu$MSM, dark matter and baryon asymmetry of the universe},'' \href{http://dx.doi.org/10.1016/j.physletb.2005.06.020}{{\em Phys. Lett. B} {\bfseries 620} (2005) 17--26}, \href{http://arxiv.org/abs/hep-ph/0505013}{[{\ttfamily hep-ph/0505013}]}.

\bibitem{Canetti:2012kh}
L.~Canetti, M.~Drewes, T.~Frossard, and M.~Shaposhnikov, ``{Dark Matter, Baryogenesis and Neutrino Oscillations from Right Handed Neutrinos},'' \href{http://dx.doi.org/10.1103/PhysRevD.87.093006}{{\em Phys. Rev. D} {\bfseries 87} (2013) 093006}, \href{http://arxiv.org/abs/1208.4607}{[{\ttfamily 1208.4607}]}.

\bibitem{Roy:2010xq}
A.~Roy and M.~Shaposhnikov, ``{Resonant production of the sterile neutrino dark matter and fine-tunings in the [nu]MSM},'' \href{http://dx.doi.org/10.1103/PhysRevD.82.056014}{{\em Phys. Rev. D} {\bfseries 82} (2010) 056014}, \href{http://arxiv.org/abs/1006.4008}{[{\ttfamily 1006.4008}]}.

\bibitem{Liu:1993tg}
J.~Liu and G.~Segre, ``{Reexamination of generation of baryon and lepton number asymmetries by heavy particle decay},'' \href{http://dx.doi.org/10.1103/PhysRevD.48.4609}{{\em Phys. Rev. D} {\bfseries 48} (1993) 4609--4612}, \href{http://arxiv.org/abs/hep-ph/9304241}{[{\ttfamily hep-ph/9304241}]}.

\bibitem{Flanz:1994yx}
M.~Flanz, E.~A. Paschos, and U.~Sarkar, ``{Baryogenesis from a lepton asymmetric universe},'' \href{http://dx.doi.org/10.1016/0370-2693(94)01555-Q}{{\em Phys. Lett. B} {\bfseries 345} (1995) 248--252}, \href{http://arxiv.org/abs/hep-ph/9411366}{[{\ttfamily hep-ph/9411366}]}. [Erratum: Phys.Lett.B 384, 487--487 (1996), Erratum: Phys.Lett.B 382, 447--447 (1996)].

\bibitem{Flanz:1996fb}
M.~Flanz, E.~A. Paschos, U.~Sarkar, and J.~Weiss, ``{Baryogenesis through mixing of heavy Majorana neutrinos},'' \href{http://dx.doi.org/10.1016/S0370-2693(96)01337-8}{{\em Phys. Lett. B} {\bfseries 389} (1996) 693--699}, \href{http://arxiv.org/abs/hep-ph/9607310}{[{\ttfamily hep-ph/9607310}]}.

\bibitem{Covi:1996fm}
L.~Covi and E.~Roulet, ``{Baryogenesis from mixed particle decays},'' \href{http://dx.doi.org/10.1016/S0370-2693(97)00287-6}{{\em Phys. Lett. B} {\bfseries 399} (1997) 113--118}, \href{http://arxiv.org/abs/hep-ph/9611425}{[{\ttfamily hep-ph/9611425}]}.

\bibitem{Pilaftsis:1997jf}
A.~Pilaftsis, ``{CP violation and baryogenesis due to heavy Majorana neutrinos},'' \href{http://dx.doi.org/10.1103/PhysRevD.56.5431}{{\em Phys. Rev. D} {\bfseries 56} (1997) 5431--5451}, \href{http://arxiv.org/abs/hep-ph/9707235}{[{\ttfamily hep-ph/9707235}]}.

\bibitem{Pilaftsis:2003gt}
A.~Pilaftsis and T.~E.~J. Underwood, ``{Resonant leptogenesis},'' \href{http://dx.doi.org/10.1016/j.nuclphysb.2004.05.029}{{\em Nucl. Phys. B} {\bfseries 692} (2004) 303--345}, \href{http://arxiv.org/abs/hep-ph/0309342}{[{\ttfamily hep-ph/0309342}]}.

\bibitem{Alwall:2011uj}
J.~Alwall, M.~Herquet, F.~Maltoni, O.~Mattelaer, and T.~Stelzer, ``{MadGraph 5 : Going Beyond},'' \href{http://dx.doi.org/10.1007/JHEP06(2011)128}{{\em JHEP} {\bfseries 06} (2011) 128}, \href{http://arxiv.org/abs/1106.0522}{[{\ttfamily 1106.0522}]}.

\bibitem{Bierlich:2022pfr}
C.~Bierlich {\em et~al.}, ``{A comprehensive guide to the physics and usage of PYTHIA 8.3},'' \href{http://dx.doi.org/10.21468/SciPostPhysCodeb.8}{{\em SciPost Phys. Codeb.} {\bfseries 2022} (2022) 8}, \href{http://arxiv.org/abs/2203.11601}{[{\ttfamily 2203.11601}]}.

\bibitem{deFavereau:2013fsa}
{\bfseries DELPHES 3} Collaboration, J.~de~Favereau, C.~Delaere, P.~Demin, A.~Giammanco, V.~Lema{\^\i}tre, A.~Mertens, and M.~Selvaggi, ``{DELPHES 3, A modular framework for fast simulation of a generic collider experiment},'' \href{http://dx.doi.org/10.1007/JHEP02(2014)057}{{\em JHEP} {\bfseries 02} (2014) 057}, \href{http://arxiv.org/abs/1307.6346}{[{\ttfamily 1307.6346}]}.

\bibitem{NNPDF:2014otw}
{\bfseries NNPDF} Collaboration, R.~D. Ball {\em et~al.}, ``{Parton distributions for the LHC Run II},'' \href{http://dx.doi.org/10.1007/JHEP04(2015)040}{{\em JHEP} {\bfseries 04} (2015) 040}, \href{http://arxiv.org/abs/1410.8849}{[{\ttfamily 1410.8849}]}.

\bibitem{Anamiati:2016uxp}
G.~Anamiati, M.~Hirsch, and E.~Nardi, ``{Quasi-Dirac neutrinos at the LHC},'' \href{http://dx.doi.org/10.1007/JHEP10(2016)010}{{\em JHEP} {\bfseries 10} (2016) 010}, \href{http://arxiv.org/abs/1607.05641}{[{\ttfamily 1607.05641}]}.

\bibitem{Das:2017hmg}
A.~Das, P.~S.~B. Dev, and R.~N. Mohapatra, ``{Same Sign versus Opposite Sign Dileptons as a Probe of Low Scale Seesaw Mechanisms},'' \href{http://dx.doi.org/10.1103/PhysRevD.97.015018}{{\em Phys. Rev. D} {\bfseries 97} no.~1, (2018) 015018}, \href{http://arxiv.org/abs/1709.06553}{[{\ttfamily 1709.06553}]}.

\bibitem{Chen:2007fv}
M.-C. Chen, ``{TASI 2006 Lectures on Leptogenesis},'' in {\em {Theoretical Advanced Study Institute in Elementary Particle Physics}: {Exploring New Frontiers Using Colliders and Neutrinos}}, pp.~123--176.
\newblock 3, 2007.
\newblock \href{http://arxiv.org/abs/hep-ph/0703087}{[{\ttfamily hep-ph/0703087}]}.

\bibitem{Esteban:2024eli}
I.~Esteban, M.~C. Gonzalez-Garcia, M.~Maltoni, I.~Martinez-Soler, J.~P. Pinheiro, and T.~Schwetz, ``{NuFit-6.0: updated global analysis of three-flavor neutrino oscillations},'' \href{http://dx.doi.org/10.1007/JHEP12(2024)216}{{\em JHEP} {\bfseries 12} (2024) 216}, \href{http://arxiv.org/abs/2410.05380}{[{\ttfamily 2410.05380}]}.

\bibitem{T2K:2023smv}
{\bfseries T2K} Collaboration, K.~Abe {\em et~al.}, ``{Measurements of neutrino oscillation parameters from the T2K experiment using $3.6\times 10^{21}$ protons on target},'' \href{http://dx.doi.org/10.1140/epjc/s10052-023-11819-x}{{\em Eur. Phys. J. C} {\bfseries 83} no.~9, (2023) 782}, \href{http://arxiv.org/abs/2303.03222}{[{\ttfamily 2303.03222}]}.

\bibitem{NOvA:2021nfi}
{\bfseries NOvA} Collaboration, M.~A. Acero {\em et~al.}, ``{Improved measurement of neutrino oscillation parameters by the NOvA experiment},'' \href{http://dx.doi.org/10.1103/PhysRevD.106.032004}{{\em Phys. Rev. D} {\bfseries 106} no.~3, (2022) 032004}, \href{http://arxiv.org/abs/2108.08219}{[{\ttfamily 2108.08219}]}.

\bibitem{T2K:2025wet}
{\bfseries T2K, NOvA} Collaboration, S.~Abubakar {\em et~al.}, ``{Joint neutrino oscillation analysis from the T2K and NOvA experiments},'' \href{http://dx.doi.org/10.1038/s41586-025-09599-3}{{\em Nature} {\bfseries 646} no.~8086, (2025) 818--824}, \href{http://arxiv.org/abs/2510.19888}{[{\ttfamily 2510.19888}]}.

\bibitem{Klaric:2021cpi}
J.~Klari{\'c}, M.~Shaposhnikov, and I.~Timiryasov, ``{Reconciling resonant leptogenesis and baryogenesis via neutrino oscillations},'' \href{http://dx.doi.org/10.1103/PhysRevD.104.055010}{{\em Phys. Rev. D} {\bfseries 104} no.~5, (2021) 055010}, \href{http://arxiv.org/abs/2103.16545}{[{\ttfamily 2103.16545}]}.

\bibitem{Ibarra:2010xw}
A.~Ibarra, E.~Molinaro, and S.~T. Petcov, ``{TeV Scale See-Saw Mechanisms of Neutrino Mass Generation, the Majorana Nature of the Heavy Singlet Neutrinos and $(\beta\beta)_{0\nu}$-Decay},'' \href{http://dx.doi.org/10.1007/JHEP09(2010)108}{{\em JHEP} {\bfseries 09} (2010) 108}, \href{http://arxiv.org/abs/1007.2378}{[{\ttfamily 1007.2378}]}.

\bibitem{Ibarra:2011xn}
A.~Ibarra, E.~Molinaro, and S.~T. Petcov, ``{Low Energy Signatures of the TeV Scale See-Saw Mechanism},'' \href{http://dx.doi.org/10.1103/PhysRevD.84.013005}{{\em Phys. Rev. D} {\bfseries 84} (2011) 013005}, \href{http://arxiv.org/abs/1103.6217}{[{\ttfamily 1103.6217}]}.

\bibitem{Dinh:2012bp}
D.~N. Dinh, A.~Ibarra, E.~Molinaro, and S.~T. Petcov, ``{The $\mu - e$ Conversion in Nuclei, $\mu \to e \gamma, \mu \to 3e$ Decays and TeV Scale See-Saw Scenarios of Neutrino Mass Generation},'' \href{http://dx.doi.org/10.1007/JHEP08(2012)125}{{\em JHEP} {\bfseries 08} (2012) 125}, \href{http://arxiv.org/abs/1205.4671}{[{\ttfamily 1205.4671}]}. [Erratum: JHEP 09, 023 (2013)].

\bibitem{Das:2017nvm}
A.~Das and N.~Okada, ``{Bounds on heavy Majorana neutrinos in type-I seesaw and implications for collider searches},'' \href{http://dx.doi.org/10.1016/j.physletb.2017.09.042}{{\em Phys. Lett. B} {\bfseries 774} (2017) 32--40}, \href{http://arxiv.org/abs/1702.04668}{[{\ttfamily 1702.04668}]}.

\bibitem{Biedermann:2017bss}
B.~Biedermann, A.~Denner, and M.~Pellen, ``{Complete NLO corrections to W$^{+}$W$^{+}$ scattering and its irreducible background at the LHC},'' \href{http://dx.doi.org/10.1007/JHEP10(2017)124}{{\em JHEP} {\bfseries 10} (2017) 124}, \href{http://arxiv.org/abs/1708.00268}{[{\ttfamily 1708.00268}]}.

\bibitem{CMS:2012feb}
{\bfseries CMS} Collaboration, S.~Chatrchyan {\em et~al.}, ``{Identification of b-Quark Jets with the CMS Experiment},'' \href{http://dx.doi.org/10.1088/1748-0221/8/04/P04013}{{\em JINST} {\bfseries 8} (2013) P04013}, \href{http://arxiv.org/abs/1211.4462}{[{\ttfamily 1211.4462}]}.

\bibitem{CMS:2018ubm}
{\bfseries CMS} Collaboration, A.~M. Sirunyan {\em et~al.}, ``{Search for top quark partners with charge 5/3 in the same-sign dilepton and single-lepton final states in proton-proton collisions at $ \sqrt{s}=13 $ TeV},'' \href{http://dx.doi.org/10.1007/JHEP03(2019)082}{{\em JHEP} {\bfseries 03} (2019) 082}, \href{http://arxiv.org/abs/1810.03188}{[{\ttfamily 1810.03188}]}.

\bibitem{ATLAS:2018ceg}
{\bfseries ATLAS} Collaboration, M.~Aaboud {\em et~al.}, ``{Search for doubly charged scalar bosons decaying into same-sign $W$ boson pairs with the ATLAS detector},'' \href{http://dx.doi.org/10.1140/epjc/s10052-018-6500-y}{{\em Eur. Phys. J. C} {\bfseries 79} no.~1, (2019) 58}, \href{http://arxiv.org/abs/1808.01899}{[{\ttfamily 1808.01899}]}.

\bibitem{Muselli:2015kba}
C.~Muselli, M.~Bonvini, S.~Forte, S.~Marzani, and G.~Ridolfi, ``{Top Quark Pair Production beyond NNLO},'' \href{http://dx.doi.org/10.1007/JHEP08(2015)076}{{\em JHEP} {\bfseries 08} (2015) 076}, \href{http://arxiv.org/abs/1505.02006}{[{\ttfamily 1505.02006}]}.

\bibitem{Campbell:2011bn}
J.~M. Campbell, R.~K. Ellis, and C.~Williams, ``{Vector Boson Pair Production at the LHC},'' \href{http://dx.doi.org/10.1007/JHEP07(2011)018}{{\em JHEP} {\bfseries 07} (2011) 018}, \href{http://arxiv.org/abs/1105.0020}{[{\ttfamily 1105.0020}]}.

\bibitem{ATLAS:2015gtp}
{\bfseries ATLAS} Collaboration, G.~Aad {\em et~al.}, ``{Search for heavy Majorana neutrinos with the ATLAS detector in pp collisions at $ \sqrt{s}=8 $ TeV},'' \href{http://dx.doi.org/10.1007/JHEP07(2015)162}{{\em JHEP} {\bfseries 07} (2015) 162}, \href{http://arxiv.org/abs/1506.06020}{[{\ttfamily 1506.06020}]}.

\bibitem{CMS:2015qur}
{\bfseries CMS} Collaboration, V.~Khachatryan {\em et~al.}, ``{Search for heavy Majorana neutrinos in $\mu^\pm \mu^\pm+$ jets events in proton-proton collisions at $\sqrt{s}$ = 8 TeV},'' \href{http://dx.doi.org/10.1016/j.physletb.2015.06.070}{{\em Phys. Lett. B} {\bfseries 748} (2015) 144--166}, \href{http://arxiv.org/abs/1501.05566}{[{\ttfamily 1501.05566}]}.

\bibitem{CMS:2018jxx}
{\bfseries CMS} Collaboration, A.~M. Sirunyan {\em et~al.}, ``{Search for heavy Majorana neutrinos in same-sign dilepton channels in proton-proton collisions at $ \sqrt{s}=13 $ TeV},'' \href{http://dx.doi.org/10.1007/JHEP01(2019)122}{{\em JHEP} {\bfseries 01} (2019) 122}, \href{http://arxiv.org/abs/1806.10905}{[{\ttfamily 1806.10905}]}.

\bibitem{ATLAS:2019kpx}
{\bfseries ATLAS} Collaboration, G.~Aad {\em et~al.}, ``{Search for heavy neutral leptons in decays of $W$ bosons produced in 13 TeV $pp$ collisions using prompt and displaced signatures with the ATLAS detector},'' \href{http://dx.doi.org/10.1007/JHEP10(2019)265}{{\em JHEP} {\bfseries 10} (2019) 265}, \href{http://arxiv.org/abs/1905.09787}{[{\ttfamily 1905.09787}]}.

\bibitem{CMS:2018iaf}
{\bfseries CMS} Collaboration, A.~M. Sirunyan {\em et~al.}, ``{Search for heavy neutral leptons in events with three charged leptons in proton-proton collisions at $\sqrt{s} =$ 13 TeV},'' \href{http://dx.doi.org/10.1103/PhysRevLett.120.221801}{{\em Phys. Rev. Lett.} {\bfseries 120} no.~22, (2018) 221801}, \href{http://arxiv.org/abs/1802.02965}{[{\ttfamily 1802.02965}]}.

\bibitem{L3:2001zfe}
{\bfseries L3} Collaboration, P.~Achard {\em et~al.}, ``{Search for heavy isosinglet neutrino in $e^{+} e^{-}$ annihilation at LEP},'' \href{http://dx.doi.org/10.1016/S0370-2693(01)00993-5}{{\em Phys. Lett. B} {\bfseries 517} (2001) 67--74}, \href{http://arxiv.org/abs/hep-ex/0107014}{[{\ttfamily hep-ex/0107014}]}.

\bibitem{DELPHI:1996qcc}
{\bfseries DELPHI} Collaboration, P.~Abreu {\em et~al.}, ``{Search for neutral heavy leptons produced in Z decays},'' \href{http://dx.doi.org/10.1007/s002880050370}{{\em Z. Phys. C} {\bfseries 74} (1997) 57--71}. [Erratum: Z.Phys.C 75, 580 (1997)].

\bibitem{CMS:2024bni}
{\bfseries CMS} Collaboration, A.~Hayrapetyan {\em et~al.}, ``{Review of searches for vector-like quarks, vector-like leptons, and heavy neutral leptons in proton{\textendash}proton collisions at {\ensuremath{\sqrt{}}}s=13 TeV at the CMS experiment},'' \href{http://dx.doi.org/10.1016/j.physrep.2024.09.012}{{\em Phys. Rept.} {\bfseries 1115} (2025) 570--677}, \href{http://arxiv.org/abs/2405.17605}{[{\ttfamily 2405.17605}]}.

\bibitem{delAguila:2008pw}
F.~del Aguila, J.~de~Blas, and M.~Perez-Victoria, ``{Effects of new leptons in Electroweak Precision Data},'' \href{http://dx.doi.org/10.1103/PhysRevD.78.013010}{{\em Phys. Rev. D} {\bfseries 78} (2008) 013010}, \href{http://arxiv.org/abs/0803.4008}{[{\ttfamily 0803.4008}]}.

\bibitem{deBlas:2013gla}
J.~de~Blas, ``{Electroweak limits on physics beyond the Standard Model},'' \href{http://dx.doi.org/10.1051/epjconf/20136019008}{{\em EPJ Web Conf.} {\bfseries 60} (2013) 19008}, \href{http://arxiv.org/abs/1307.6173}{[{\ttfamily 1307.6173}]}.

\bibitem{Akhmedov:2013hec}
E.~Akhmedov, A.~Kartavtsev, M.~Lindner, L.~Michaels, and J.~Smirnov, ``{Improving Electro-Weak Fits with TeV-scale Sterile Neutrinos},'' \href{http://dx.doi.org/10.1007/JHEP05(2013)081}{{\em JHEP} {\bfseries 05} (2013) 081}, \href{http://arxiv.org/abs/1302.1872}{[{\ttfamily 1302.1872}]}.

\bibitem{Hagedorn:1963hdh}
R.~Hagedorn, {\em {Relativistic kinematics}: {a guide to the kinematic problems of high-energy physics}}.
\newblock Benjamin, New York, NY, 1963.

\bibitem{Nachtmann:1990ta}
O.~Nachtmann, {\em {Elementary Particle Physics: Concepts and Phenomena}}.
\newblock 1990.

\bibitem{Byckling:1971vca}
E.~Byckling and K.~Kajantie, {\em {Particle Kinematics}: {(Chapters I-VI, X)}}.
\newblock University of Jyvaskyla, Jyvaskyla, Finland, 1971.

\bibitem{Chun:2019nwi}
E.~J. Chun, A.~Das, S.~Mandal, M.~Mitra, and N.~Sinha, ``{Sensitivity of Lepton Number Violating Meson Decays in Different Experiments},'' \href{http://dx.doi.org/10.1103/PhysRevD.100.095022}{{\em Phys. Rev. D} {\bfseries 100} no.~9, (2019) 095022}, \href{http://arxiv.org/abs/1908.09562}{[{\ttfamily 1908.09562}]}.

\bibitem{A:2025ygb}
S.~K. A., S.~Das, A.~Das, and S.~Mandal, ``{Right-handed neutrino production from Z' interactions in forward search experiments},'' \href{http://dx.doi.org/10.1103/s9jw-ckzm}{{\em Phys. Rev. D} {\bfseries 112} no.~11, (2025) 115045}, \href{http://arxiv.org/abs/2508.10734}{[{\ttfamily 2508.10734}]}.

\bibitem{ParticleDataGroup:2010dbb}
{\bfseries Particle Data Group} Collaboration, K.~Nakamura {\em et~al.}, ``{Review of particle physics},'' \href{http://dx.doi.org/10.1088/0954-3899/37/7A/075021}{{\em J. Phys. G} {\bfseries 37} (2010) 075021}.

\bibitem{CLEO:2005jsh}
{\bfseries CLEO} Collaboration, M.~Artuso {\em et~al.}, ``{Improved measurement of B (D+ ---{\ensuremath{>}} mu+ nu) and the pseudoscalar decay constant f(D+)},'' \href{http://dx.doi.org/10.1103/PhysRevLett.95.251801}{{\em Phys. Rev. Lett.} {\bfseries 95} (2005) 251801}, \href{http://arxiv.org/abs/hep-ex/0508057}{[{\ttfamily hep-ex/0508057}]}.

\bibitem{MILC:2002lnl}
{\bfseries MILC} Collaboration, C.~Bernard, S.~Datta, T.~DeGrand, C.~DeTar, S.~Gottlieb, U.~M. Heller, C.~McNeile, K.~Orginos, R.~Sugar, and D.~Toussaint, ``{Lattice calculation of heavy light decay constants with two flavors of dynamical quarks},'' \href{http://dx.doi.org/10.1103/PhysRevD.66.094501}{{\em Phys. Rev. D} {\bfseries 66} (2002) 094501}, \href{http://arxiv.org/abs/hep-lat/0206016}{[{\ttfamily hep-lat/0206016}]}.

\bibitem{Ivanov:2006ni}
M.~A. Ivanov, J.~G. Korner, and P.~Santorelli, ``{Exclusive semileptonic and nonleptonic decays of the $B_c$ meson},'' \href{http://dx.doi.org/10.1103/PhysRevD.73.054024}{{\em Phys. Rev. D} {\bfseries 73} (2006) 054024}, \href{http://arxiv.org/abs/hep-ph/0602050}{[{\ttfamily hep-ph/0602050}]}.

\bibitem{Feldmann:1999uf}
T.~Feldmann, ``{Quark structure of pseudoscalar mesons},'' \href{http://dx.doi.org/10.1142/S0217751X00000082}{{\em Int. J. Mod. Phys. A} {\bfseries 15} (2000) 159--207}, \href{http://arxiv.org/abs/hep-ph/9907491}{[{\ttfamily hep-ph/9907491}]}.

\bibitem{Ebert:2006hj}
D.~Ebert, R.~N. Faustov, and V.~O. Galkin, ``{Relativistic treatment of the decay constants of light and heavy mesons},'' \href{http://dx.doi.org/10.1016/j.physletb.2006.02.042}{{\em Phys. Lett. B} {\bfseries 635} (2006) 93--99}, \href{http://arxiv.org/abs/hep-ph/0602110}{[{\ttfamily hep-ph/0602110}]}.

\end{thebibliography}\endgroup
\end{document}